\documentclass[aps,prd,showkeys,superscriptaddress,singlecolumn,nofootinbib,floatfix]{revtex4}
\usepackage{amsmath}
\usepackage{amssymb}
\usepackage{amsthm}
\usepackage{mathrsfs}
\usepackage{graphicx}
\usepackage{epstopdf}
\usepackage{fancyhdr}
\usepackage{array}
\usepackage[all]{xy}
\usepackage{eufrak}
\usepackage{euscript}
\usepackage{enumerate}
\usepackage{slashed}
\usepackage{hyperref}
\usepackage{subfigure}
\usepackage{epstopdf}

\pdfoutput=1
\hypersetup{pdftex,colorlinks=true,linkcolor=blue,citecolor=blue,menucolor=black,urlcolor=blue,filecolor=blue}

\begin{document}

\title{Hydrodynamic attractor and the fate of perturbative expansions in
Gubser flow}
\date{\today}
\author{Gabriel S.~Denicol}
\affiliation{Instituto de F\'isica, Universidade Federal Fluminense, UFF, Niter\'oi,
24210-346, RJ, Brazil}
\author{Jorge Noronha}
\affiliation{Instituto de F\'isica, Universidade de S\~ao Paulo, Rua do Mat\~ao, 1371,
Butant\~a, 05508-090, S\~ao Paulo, SP, Brazil}

\begin{abstract}
Perturbative expansions, such as the well-known gradient series and the
recently proposed slow-roll expansion, have been recently used to
investigate the emergence of hydrodynamic behavior in systems undergoing
Bjorken flow. In this paper we determine for the first time the large order
behavior of these perturbative expansions in relativistic hydrodynamics in
the case of Gubser flow. While both series diverge, the slow-roll series can
provide a much better overall description of the system's dynamics than the
gradient expansion when both series are truncated at low orders. The
truncated slow-roll series can also describe the attractor solution of
Gubser flow as long as the system is sufficiently close to equilibrium near
the origin (i.e., $\rho=0$) in $dS_3 \otimes \mathbb{R}$. Differently than
the case of Bjorken flow, here we show that the Gubser flow attractor
solution is not solely a function of the effective Knudsen number $\tau_R 
\sqrt{\sigma_{\mu\nu}\sigma^{\mu\nu}} \sim \tau_R\, \tanh\rho$. Our results
give further support to the idea that new \emph{resummed} constitutive
relations between dissipative currents and the gradients of conserved
quantities can emerge in systems far from equilibrium that are beyond the
regime of validity of the usual gradient expansion.
\end{abstract}

\maketitle
\tableofcontents



\section{Introduction}

\label{sec:intro}

Hydrodynamic behavior in a many-body system is usually characterized by the
existence of constitutive relations between dissipative currents and
gradients of conserved quantities \cite{chapman-cowling}. When the gradients
are small compared to the system's corresponding microscopic scales, i.e.,
when the Knudsen number is small, the system is close to local equilibrium
and, in principle, dissipative corrections can be taken into account via a
systematic expansion in powers of the Knudsen number \cite{chapman-cowling}.
Such an expansion, when truncated to first order, leads to the famous
non-relativistic Navier-Stokes (NS) equations \cite{navier,stokes}.

The relativistic generalization of these equations was worked out by Eckart 
\cite{Eckart:1940te} and Landau \cite{landau} nearly a century ago and, with
the advent of the quark-gluon plasma formed in ultrarelativistic heavy ion
collisions, viscous relativistic hydrodynamics has become the main tool to
describe the spacetime evolution of the hot and dense matter created in
these collisions \cite{Heinz:2013th,deSouza:2015ena}. This led to a number
of new theoretical approaches, e.g. \cite%
{Koide:2006ef,Baier:2007ix,Bhattacharyya:2008jc,Denicol:2010xn,Florkowski:2010cf,Martinez:2010sc,Denicol:2012cn,Florkowski:2013lza,Denicol:2014xca,Denicol:2014tha,Heller:2015dha,Bemfica:2017wps,Blaizot:2017ucy}
as well as computational/phenomenological developments \cite%
{Romatschke:2007mq,Song:2007ux,Luzum:2008cw,Denicol:2009am,Schenke:2010rr,Gale:2012rq,DelZanna:2013eua,Noronha-Hostler:2013gga,Noronha-Hostler:2014dqa,Shen:2014vra,Ryu:2015vwa,Bernhard:2016tnd,Bazow:2016yra,Okamoto:2017ukz,Pang:2018lqw}
in viscous relativistic hydrodynamics, which have built upon Israel and
Stewart's seminal work \cite{Israel:1976tn,Stewart:1977,Israel:1979wp} and
extended it in a number of ways (for a recent review, see \cite%
{Romatschke:2017ejr}).

However, the extreme energy density and large spatial gradients expected to
occur in the initial stages of the quark-gluon plasma formed in the
collisions of large nuclei \cite{Gale:2012rq}, together with the later
measurement of large collective behavior also in small collision systems 
\cite{Khachatryan:2015waa}, have contributed in part to the question of
whether hydrodynamic behavior can also appear when gradients are not small
and more terms in the gradient series have to be taken into account.
Reference\ \cite{Heller:2013fn} initiated the study of the large order
behavior of the gradient series in the field, which was shown to have zero
radius of convergence in the case of a strongly coupled $\mathcal{N}=4$
supersymmetric Yang-Mills plasma undergoing Bjorken expansion \cite{bjorken}%
. Other examples later followed involving strongly coupled systems with
different symmetries in the context of cosmology \cite{Buchel:2016cbj} and
also in kinetic theory \cite{Denicol:2016bjh,Heller:2016rtz} (for a review,
see \cite{Florkowski:2017olj}). To extract the properties and meaningfully
handle these divergent gradient series, powerful mathematical techniques
from resurgence theory (see, e.g.\ \cite{Aniceto:2018bis}) have been used to
resum the large order behavior of the series in systems with large degree of
symmetry \cite%
{Heller:2015dha,Basar:2015ava,Aniceto:2015mto,Buchel:2016cbj,Florkowski:2016zsi,Heller:2016rtz,Heller:2018qvh}%
.

Now that it is known that the gradient series diverges and that it is still
meaningful to look for hydrodynamic behavior even in situations where
gradients are not necessarily small, one may ask if a different mathematical
representation, which does not rely on the assumption of small gradients,\
can be formulated to describe the hydrodynamic regime. The first step in
this direction was made in \cite{Heller:2015dha} with the proposal that
hydrodynamic behavior may be meaningfully defined even far-from-equilibrium
as long as a late time \textquotedblleft attractor" structure is present. In
this context, NS already played the role of an attractor since the system,
regardless of its initial conditions, approaches this limit when
sufficiently close to equilibrium - the novelty of Ref.\ \cite%
{Heller:2015dha}'s proposal is that this may occur in the far from
equilibrium regime as well. Several works have since then investigated this
in Bjorken flow \cite%
{Romatschke:2017vte,Spalinski:2017mel,Strickland:2017kux,Romatschke:2017acs,Florkowski:2017jnz,Denicol:2017lxn,Casalderrey-Solana:2017zyh,Almaalol:2018ynx}
but the question of what happens in systems with less symmetries still
remains \cite{Romatschke:2017acs}.

In Bjorken flow, an approximation to the attractor solution was obtained 
\cite{Heller:2015dha} using a method analogous to the slow-roll expansion
developed in cosmology \cite{Liddle:1994dx}. In the context of
hydrodynamics, a given order in the slow-roll expansion contains derivative
terms of all orders, which suggests that such an expansion can be useful in
the formulation of hydrodynamics of far-from-equilibrium systems. The large
order behavior of the slow-roll expansion was computed for the first time in 
\cite{Denicol:2017lxn} where it was shown that this series also diverges in
Bjorken flow. Since this type of perturbative expansion is fairly recent in
hydrodynamic applications, it is useful to check different types of systems
and flow profiles where the slow-roll expansion can be systematically
implemented to obtain a better understanding of its properties.

Divergent series are known to provide excellent approximations to the
solution of several mathematical problems \cite{bender} and, thus, the fact
that both series diverge is not an issue per se. The relevant question is
which one of these divergent series provides the best approximation to the
out-of-equilibrium dynamics of the system under consideration after
truncation. The detailed analysis performed in \cite{Denicol:2017lxn} for
Bjorken flow suggested that the slow-roll expansion may lead to a better
overall description of the system's dynamics for a wider range of values of
Knudsen number in comparison to the gradient series.

In this work we investigate the fate of both series in a simple system
undergoing Gubser flow \cite{Gubser}. This type of flow describes a
relativistic fluid that is not only longitudinally boost invariant but it
also undergoes a radially symmetric expansion in the transverse plane. For
simplicity, we follow \cite{Heller:2015dha} and consider as our
\textquotedblleft microscopic" theory the Israel-Stewart formulation of
transient fluid dynamics \cite{Israel:1976tn,Stewart:1977,Israel:1979wp},
whose properties in Gubser flow were first studied in \cite%
{Marrochio:2013wla}. We explain how to systematically construct the gradient
and slow-roll perturbative series in this type of flow and we compute the
large order behavior of both series for the first time in Gubser flow. Our
results strongly indicate that both series are divergent, as happened for
systems expanding following Bjorken flow. However, when comparing low order
truncations of both series, we observe that the slow-roll expansion can
provide a better overall description of exact solutions in a wider range of
parameters in comparison to the gradient expansion. The truncated slow-roll
series can also describe the attractor solution of Gubser flow as long as
the shear stress tensor approximately vanishes near the origin of the
time-like coordinate (i.e., $\rho =0$) in $dS_{3}\otimes \mathbb{R}$.
Differently than the case of Bjorken flow, we find that the Gubser flow
attractor solution is not solely a function of the effective Knudsen number $%
\tau _{R}\sqrt{\sigma _{\mu \nu }\sigma ^{\mu \nu }}\sim \tau _{R}\,\tanh
\rho $. Our results support the idea that a new type of constitutive
relations between dissipative currents (e.g. the shear stress tensor) and
the gradients of conserved quantities can emerge in far-from-equilibrium
systems which are, thus, beyond the regime of applicability of the gradient
expansion.

This paper is organized as follows. In the next section we define the
equations of motion of the Israel-Stewart-like theory considered in this
paper while in Sec.\ \ref{sec:gubser} we define what is Gubser flow. Sec.\ %
\ref{gradientscheme1} and \ref{gradientscheme2} are devoted to explain how
to perform the same gradient series in two different ways while in Sec.\ \ref%
{sec:coldplasma} we discuss a complementary perturbative series in powers of
the inverse relaxation time. We develop the slow-roll expansion in Sec.\ \ref%
{sec:slowroll} and investigate the hydrodynamic attractor of Gubser flow in
Sec.\ \ref{sec:attractor}. Our final remarks can be found in Sec.\ \ref%
{sec:conclusions}. Appendix \ref{sec:appendixA} gives yet another way to
develop the gradient series in Gubser flow.

\emph{Definitions}: We use a mostly minus metric signature and natural
units, $\hbar = c = k_B = 1$.

\section{Conformal Israel-Stewart Theory}

\label{sec:hydro}

Excluding the contribution from conserved charges, the main fluid-dynamical
equations are the continuity equations related to the conservation of energy
and momentum 
\begin{equation}
\nabla _{\mu }T^{\mu \nu }=0.
\end{equation}%
The field $T^{\mu \nu }$ introduced above is the energy-momentum tensor,
which is commonly decomposed in terms of the local fluid velocity $u^{\mu }$
as 
\begin{equation}
T^{\mu \nu }=\varepsilon u^{\mu }u^{\nu }-\Delta ^{\mu \nu }P+\pi ^{\mu \nu
},
\end{equation}%
where $\varepsilon $ is the energy density, $P$ is the thermodynamic
pressure and $\pi ^{\mu \nu }$ is the shear stress tensor. The fluid
velocity is constructed to be a normalized 4-vector, $u_{\mu }u^{\mu }=1$,
defined according to Landau's picture \cite{landau}, $T^{\mu \nu }u_{\nu
}\equiv \varepsilon u^{\mu }$, as an eigenvector of the energy-momentum
tensor. Note that we also introduced the projection operator transverse to $%
u^{\mu }$, $\Delta _{\mu \nu }\equiv g_{\mu \nu }-u_{\mu }u_{\nu }$, with $%
g_{\mu \nu }$ being the spacetime metric. In this paper, we only consider
the dynamics of conformal fluids \cite{Baier:2007ix} and, consequently,
there is no bulk viscous pressure and the equation of state of the fluid is
given by, $\varepsilon =3P$ (i.e., the trace of $T^{\mu \nu }$ vanishes).

The conservation laws alone do not describe all the degrees of freedom of $%
T^{\mu \nu }$. They must be complemented by additional dynamical equations
(or constitutive relations), that describe the time evolution of the
shear stress tensor. For this purpose, we employ the transient hydrodynamic
equations derived by Israel and Stewart \cite%
{Israel:1976tn,Stewart:1977,Israel:1979wp}\ from kinetic theory (and later
complemented by several authors \cite%
{Muronga:2001zk,Muronga:2003ta,Denicol:2011fa,Denicol:2012es,Denicol:2012vq,Molnar:2013lta,Denicol:2014vaa,Denicol:2014loa,El:2009vj,Jaiswal:2013vta}%
). In this framework, the shear stress tensor satisfies the following
relaxation-type equation 
\begin{equation}
\tau _{R}\Delta _{\alpha \beta }^{\mu \nu }D\pi ^{\alpha \beta }+\delta
_{\pi \pi }\,\Theta \,\pi ^{\mu \nu }+\tau _{\pi \pi }\,\Delta _{\alpha
\beta }^{\mu \nu }\pi ^{\alpha \lambda }\sigma _{\lambda }^{\beta }-2\,\tau
_{R}\Delta _{\alpha \beta }^{\mu \nu }\pi _{\lambda }^{\alpha }\omega
^{\beta \lambda }+\pi ^{\mu \nu }=2\eta \sigma ^{\mu \nu },
\end{equation}%
where $D=u^{\mu }\nabla _{\mu }$ is the co-moving covariant derivative, $%
\Theta =\nabla _{\mu }u^{\mu }$ is the local expansion rate, $\sigma _{\mu
\nu }=\Delta _{\mu \nu }^{\alpha \beta }\nabla _{\alpha }u_{\beta }$ is the
shear tensor with $\Delta _{\mu \nu }^{\alpha \beta }=\frac{1}{2}\left(
\Delta _{\mu }^{\alpha }\Delta _{\nu }^{\beta }+\Delta _{\nu }^{\alpha
}\Delta _{\mu }^{\beta }-\frac{2}{3}\Delta ^{\alpha \beta }\Delta _{\mu \nu
}\right) $, $\omega _{\mu \nu }=(\Delta _{\mu }^{\lambda }\nabla _{\lambda
}u_{\nu }-\Delta _{\nu }^{\lambda }\nabla _{\lambda }u_{\mu })/2$ is the
vorticity tensor, $\eta $ is the shear viscosity, and $\tau _{R}$ is the
shear relaxation time.

The equations above may be derived from the Boltzmann equation using the
14-moment approximation or the relaxation time approximation (RTA), as shown
in Refs.\ \cite{Denicol:2012cn,Denicol:2012es,Denicol:2014vaa}. For a
massless gas in the 14-moment approximation, it was demonstrated that $%
\delta _{\pi \pi }=4/3\,\tau _{R}$, $\tau _{\pi \pi }=10/21\,\tau _{R}$ and $%
\eta =(\varepsilon +P)\tau _{R}/5$. Since we are dealing with a conformal
fluid, the shear relaxation time must be inversely proportional to the
temperature 
\begin{equation}
\tau _{R}=\frac{c}{T},
\end{equation}%
where $c$ is a constant that will determine the magnitude of the shear
viscosity to entropy density ratio, $\eta /s=c/5$. For the sake of
simplicity, we neglect the transport coefficient $\tau _{\pi \pi }$ in this
work. Furthermore, the nonlinear term that contains the vorticity vanishes
in the flow profile investigated in this paper (see next section) and does
not have to be considered as well. This leads to the so-called (simplified)
conformal Israel-Stewart equations \cite{Marrochio:2013wla} 
\begin{equation}
\Delta _{\alpha \beta }^{\mu \nu }D\pi ^{\alpha \beta }+\frac{4}{3}\,\pi
^{\mu \nu }\Theta +\frac{\pi ^{\mu \nu }}{\tau _{R}}=2\frac{\eta }{\tau _{R}}%
\sigma ^{\mu \nu }.
\end{equation}

\section{Gubser Flow}

\label{sec:gubser}

Following \cite{Gubser}, we look for solutions of the hydrodynamic equations
with $SO(3)\otimes SU(1,1)\otimes Z_{2}$ symmetry in flat spacetime. This is
more naturally implemented by performing a Weyl transformation to $%
dS_{3}\otimes \mathbb{R}$ spacetime (where $dS_{3}$ stands for the
3-dimensional de Sitter spacetime \cite{Hawking:1973uf}) and assuming that
the fluid is homogeneous in this curved geometry. This spacetime is
described by the line element \cite{Gubser}%
\begin{equation}
ds^{2}=g_{\mu \nu }dx^{\mu }dx^{\nu }=d\rho ^{2}-\left( \cosh ^{2}\rho
d\theta ^{2}+\cosh ^{2}\rho \sin ^{2}\theta d\phi ^{2}+d\eta ^{2}\right) \,.
\label{gubserlineelement}
\end{equation}%
The nonzero Christoffel symbols of this metric are $\Gamma _{\theta \theta
}^{\rho }=\cosh \rho \sinh \rho $, $\Gamma _{\phi \phi }^{\rho }=\left( \sin
\theta \right) ^{2}\cosh \rho \sinh \rho $, $\Gamma _{\rho \theta }^{\theta
}=\Gamma _{\rho \phi }^{\phi }=\tanh \rho $, $\Gamma _{\phi \phi }^{\theta
}=-\sin \theta \cos \theta $, $\Gamma _{\theta \phi }^{\phi }=\left( \tan
\theta \right) ^{-1}$ and its determinant is $\sqrt{-g}=\sin \theta \left(
\cosh \rho \right) ^{2}$. Since the system is homogeneous, all fields depend
only on the time-like variable $\rho $, without displaying any dependence on 
$\theta $, $\phi $, and $\eta $. If we transform back to Minkowski spacetime
in hyperbolic coordinates $(\tau ,r,\phi ,\eta )$, where the line element is 
$ds^{2}=d\tau ^{2}-dr^{2}-r^{2}d\phi ^{2}-\tau ^{2}d\eta $ \cite{Gubser},
this homogeneous system is mapped into a longitudinally boost invariant
fluid whose expansion in the transverse plane is radially symmetric. In flat
spacetime, this type of fluid displays a more complex pattern of expansion
in comparison to the Bjorken solutions \cite{bjorken} considered in previous
works, e.g. \cite%
{Heller:2013fn,Heller:2015dha,Florkowski:2016zsi,Denicol:2016bjh,Denicol:2017lxn}
where no radial expansion is allowed.

The assumption that the system is homogeneous in $dS_3 \otimes \mathbb{R}$
leads to a trivial velocity field, $u^{\mu }=\left( 1,0,0,0\right) $, which
automatically satisfies the momentum conservation continuity equations.
Nevertheless, in this curved space-time, the fluid still has a nonzero
expansion rate and shear tensor, which are given by 
\begin{eqnarray}
\Theta &\equiv &\nabla_{\mu }u^{\mu }=2\tanh \rho , \\
\sigma _{\mu \nu } &\equiv &\frac{1}{2}\Delta _{\mu }^{\alpha }\Delta _{\nu
}^{\beta }\left( \nabla_{\alpha }u_{\beta }+\nabla_{\beta }u_{\alpha
}\right) -\frac{1}{3}\Delta _{\mu \nu }\Theta \\
&=&\mathrm{diag}\left( 0,g_{\theta \theta },g_{\phi \phi },-2g_{\eta \eta
}\right) \times \frac{1}{3}\tanh \rho ,
\end{eqnarray}%
where $\nabla_{\alpha }u_{\beta }=\partial _{\alpha }u_{\beta }-\Gamma
_{\alpha \beta }^{\lambda }u_{\lambda }$ is the covariant derivative of the
flow velocity. As already stated, the vorticity tensor vanishes for this
flow profile, $\omega ^{\mu \nu }=0$, and plays no role in the results
obtained in this paper. Since the shear tensor is diagonal, the shear stress
tensor will also be diagonal, $\pi ^{\mu \nu }=\mathrm{diag}\left( 0,\pi
^{\theta \theta },\pi ^{\phi \phi },\pi ^{\eta \eta }\right) $, as long as
it is initially diagonal. The equation of motion for the energy density, $%
\varepsilon $ , then becomes%
\begin{equation*}
u_{\nu }\nabla_{\mu }T^{\mu \nu }=\frac{d\varepsilon}{d\rho} +\frac{8}{3}%
\varepsilon \tanh \rho -\frac{1}{3}\pi ^{\eta \eta }\tanh \rho =0,
\end{equation*}%
where we used that $\pi ^{\mu \nu }\sigma _{\mu \nu }=\pi ^{\eta \eta }\tanh
\rho $. The equations of motion for $\pi ^{\mu \nu }$ are%
\begin{eqnarray}
\frac{d\pi _{\theta }^{\theta }}{d\rho}+\frac{8}{3}\,\pi _{\theta }^{\theta
}\tanh \rho +\frac{\pi _{\theta }^{\theta }}{\tau _{R}} &=&\frac{2\eta }{%
3\tau _{R}}\tanh \rho , \\
\frac{d\pi _{\phi }^{\phi }}{d\rho}+\frac{8}{3}\,\pi _{\phi }^{\phi }\tanh
\rho +\frac{\pi _{\phi }^{\phi }}{\tau _{R}} &=&\frac{2\eta }{3\tau _{R}}%
\tanh \rho , \\
\frac{d\pi _{\eta }^{\eta }}{d\rho}+\frac{8}{3}\pi _{\eta }^{\eta }\tanh
\rho +\frac{\pi _{\eta }^{\eta }}{\tau _{R}} &=&-\frac{4\eta }{3\tau _{R}}%
\tanh \rho .
\end{eqnarray}

For the purposes of this paper, it is convenient to re-express these
equations in terms of the temperature, $T$ (the system is conformal, $%
\varepsilon \sim T^{4}$), and a variable $\pi $ defined as%
\begin{equation}
\pi \equiv -\frac{\pi _{\eta }^{\eta }}{\varepsilon +P}.  \label{definepinew}
\end{equation}%
With this choice of variables, the simplified conformal hydrodynamical
equations of Israel-Stewart theory become \cite{Marrochio:2013wla} 
\begin{eqnarray}
\frac{1}{T}\frac{dT}{d\rho}+\frac{2}{3}\tanh \rho -\frac{1}{3}\pi \tanh \rho
&=&0,  \label{Good} \\
\frac{d\pi}{d\rho} +\frac{\pi }{\tau _{R}}+\frac{4}{3}\pi ^{2}\tanh \rho &=&%
\frac{4}{15}\tanh \rho .  \label{Great}
\end{eqnarray}%
These are the equations that we will investigated throughout this paper.

\section{Gradient Expansion}

\label{gradientscheme1}

We develop our calculations following the procedure outlined in Ref.\ \cite%
{Denicol:2017lxn}. We convert the original problem into a perturbation
theory problem by introducing a dimensionless parameter $\epsilon $ into the
differential equation satisfied by $\pi $ as follows 
\begin{equation}
\epsilon \frac{d\pi }{d\rho }+\frac{\pi T}{c}+\frac{4}{3}\pi ^{2}\epsilon
\tanh \rho =\frac{4}{15}\epsilon \tanh \rho ,
\end{equation}%
where we already used that $\tau _{R}=c/T$. Now, in this problem the
shear stress tensor and the temperature are functions of $\rho $ and the new
variable $\epsilon $, i.e., $T=T(\rho ,\epsilon )$ and $\pi =\pi (\rho
,\epsilon )$. The parameter $\epsilon $ was introduced in every term which
contains a derivative or $\tanh \rho $ and, hence, it becomes a book-keeping
parameter to count orders or powers of gradients. Thus, an expansion in
powers of $\epsilon $ will naturally lead to an expansion in powers of
gradients. Note that while in Bjorken scaling it was straightforward to
deduce that powers of gradients correspond to inverse powers of the time
coordinate $\tau $ \cite{Denicol:2017lxn}, here the situation is not that
simple and the corresponding terms are obtained from the perturbative
procedure itself.

Next, we look for a solution for $\pi $ that can be represented as a series
in powers of $\epsilon $ 
\begin{equation}
\pi \sim \sum\limits_{n=0}^{\infty }\pi _{n}\left( \rho \right) \epsilon
^{n}.  \label{Exp}
\end{equation}%
This reduces the problem to solving an infinite number of simpler equations
(in this case, algebraic equations), which are given by recurrence
relations. At the end of the calculation, one sets $\epsilon =1$ to recover
the parameters of the original problem. We note that such perturbative
procedure is only useful when the first few terms of the series contain at
least some basic properties of the exact solution, i.e., if a low order
truncation of the series can capture basic trends of the solution. This does
not necessarily mean that the series must converge. As a matter of fact, in
practice convergent series quite often do not offer useful representations
of functions since they may be slowly convergent and require a great
number of terms to provide a good approximation in a given domain \cite%
{bender}. On the other hand, truncations of divergent series are known to
provide very good approximations of certain functions (as in the case of the
error function).

Naturally, this procedure will not lead to a general solution of
Israel-Stewart theory since the equations obtained in this type of
perturbative approach do not contain any free parameter associated with the
initial condition for $\pi $ (the initial condition for the temperature
remains a free parameter, even in the perturbative problem). In this sense,
what will be obtained with this approach is just one solution for $\pi $
that cannot be adjusted to an arbitrary boundary condition. However, this
solution is expected to have physical meaning, reflecting the long-time,
slow evolution of the system when all transient, initial-state dynamics is
lost and the system enters a universal, hydrodynamical regime (assumed in
this section to be described by the gradient expansion).

We now continue the perturbative calculation, substituting the proposed
series solutions in powers of $\epsilon $ into the equations of motion for $%
\pi $. One then obtains the following result:%
\begin{equation}
c\sum_{n=0}^{\infty }\frac{d\pi _{n}}{d\rho }\epsilon
^{n+1}+T\sum_{n=0}^{\infty }\pi _{n}\epsilon ^{n}+\frac{4}{3}%
c\sum_{n=0}^{\infty }\sum_{m=0}^{\infty }\pi _{n}\pi _{m}\epsilon
^{n+m+1}\tanh \rho =\frac{4}{15}c\epsilon \tanh \rho .  \label{Main}
\end{equation}%
The problem in solving this equation, i.e., in collecting all terms that are
of the same power in $\epsilon $, is that the term $d\pi _{n}/d\rho $ also
has an $\epsilon $--dependence that must be considered when grouping the
terms. This can be taken into account by noticing that the coefficient $\pi
_{n}$ depends on $\rho $ through two different variables, $\pi _{n}=\pi
_{n}\left( \tanh \rho ,T\left( \rho \right) \right) $ and, thus, its
derivative can be mathematically re-expressed in the following way%
\begin{equation}
\frac{d\pi _{n}}{d\rho }=\left. \frac{\partial \pi _{n}}{\partial \rho }%
\right\vert _{T}+\left. \frac{\partial \pi _{n}}{\partial T}\right\vert
_{\tanh \rho }\frac{dT}{d\rho }.
\end{equation}%
Above, the derivatives are taken as if $\tanh \rho $ and $T$ were two
independent variables. It is the temperature derivative that carries the $%
\epsilon $--dependence and, thus, using Eq.\ \eqref{Good} we can re-write
Eq.\ \eqref{Main} as%
\begin{gather}
c\sum_{n=0}^{\infty }\left( \left. \frac{\partial \pi _{n}}{\partial \rho }%
\right\vert _{T}-\frac{2}{3}T\tanh \rho \left. \frac{\partial \pi _{n}}{%
\partial T}\right\vert _{\tanh \rho }\right) \epsilon ^{n+1}+\frac{c}{3}%
\tanh \rho \sum_{n=0}^{\infty }\sum_{m=0}^{\infty }\pi _{m}T\left. \frac{%
\partial \pi _{n}}{\partial T}\right\vert _{\tanh \rho }\epsilon ^{n+m+1} 
\notag \\
+T\sum_{n=0}^{\infty }\pi _{n}\epsilon ^{n}+\frac{4}{3}c\sum_{n=0}^{\infty
}\pi _{n}\pi _{m}\epsilon ^{n+m+1}\tanh \rho -\frac{4}{15}c\epsilon \tanh
\rho =0.
\end{gather}

Now that the terms are properly organized in powers of $\epsilon $ we can\
group together the terms that are of the same order and obtain the set of
recurrence relations that must be solved to obtain $\pi _{n}$. The zeroth
order term must satisfy%
\begin{equation}
T\pi _{0}=0,
\end{equation}%
which describes a system that is in local equilibrium, as expected.
Collecting the terms that are of first order in $\epsilon $ one obtains%
\begin{equation}
\pi _{1}(\rho )=\frac{4}{15}\tau _{R}\tanh \rho ,
\end{equation}%
which corresponds to relativistic NS theory. This expression also reflects
the fact that the shear stress tensor in NS theory is linear in the Knudsen
number $K_{N}\sim \tau _{R}\sqrt{\sigma _{\mu \nu }\sigma ^{\mu \nu }}$ for
Gubser flow. Finally, the terms that are of second order or higher in $%
\epsilon $, ($n\geq 2$), satisfy the equations 
\begin{eqnarray}
\frac{T}{c}\pi _{n+1} &=&-\frac{4}{3}\sum_{m=0}^{n}\pi _{n-m}\pi _{m}\tanh
\rho -\frac{1}{3}\tanh \rho \sum_{m=0}^{n}\pi _{m}T\left. \frac{\partial \pi
_{n-m}}{\partial T}\right\vert _{\tanh \rho }  \notag \\
&&+\left( \frac{2}{3}\tanh \rho \right) \,T\left. \frac{\partial \pi _{n}}{%
\partial T}\right\vert _{\tanh \rho }-\left. \frac{\partial \pi _{n}}{%
\partial \rho }\right\vert _{T}.
\end{eqnarray}

Since we already know $\pi _{0}$ and $\pi _{1}$, we can calculate all the $%
\pi _{n}$'s that follow. For the sake of completeness, we write down below
the answer up to third order 
\begin{eqnarray}
\pi &=&\frac{4}{15}\epsilon \tau _{R}\tanh \rho -\frac{4}{15}\left( \epsilon
\tau _{R}\right) ^{2}\left( 1-\frac{1}{3}\tanh ^{2}\rho \right) \nonumber \\
&&+\frac{8}{45}\left( \epsilon \tau _{R}\right) ^{3}\left( \tanh \rho -\frac{%
1}{15}\tanh ^{3}\rho \right) +\mathcal{O}\left( \epsilon ^{4}\right) .
\label{grad3rdorder}
\end{eqnarray}%
Therefore, we can write the solution as a series in powers of $\epsilon \tau
_{R}$, with all the temperature contribution to the shear stress tensor
contained in the relaxation time. Also, we note that while the NS result
depends solely on the combination $\tau _{R}\tanh \rho $, the same is not
true for the higher order terms in the gradient expansion, which depend
separately on $\tau _{R}$\ and $\tanh \rho $.

These equations have the form that is traditionally associated with a
gradient expansion: the zeroth and first order truncations obtained in this
section correspond to well known results, ideal hydrodynamics and
Navier-Stokes theory, respectively. Both of these examples were already studied
extensively in the literature \cite{Gubser}. In Appendix \ref{sec:appendixA}
we derive the gradient expansion using another equivalent method.

\section{Gradient Expansion -- Revisited}

\label{gradientscheme2}

In the previous section we constructed the gradient expansion solution of
Israel-Stewart theory following a perturbative scheme. We demonstrated that
the result can be expressed as an expansion in powers of $\epsilon \tau _{R}$%
. In this section we construct once again the gradient expansion solution of
Israel-Stewart theory, but now using the knowledge that the temperature
appears in the series only through powers of the relaxation time, $\tau
_{R}\sim 1/T$. Therefore, it is more straightforward to simply assume an
expansion of $\pi $ just in powers of $\tau _{R}$ from the very beginning
(the parameter $\epsilon $ can also be included, as before, but it will not
make any difference),%
\begin{equation}
\pi =\sum_{n=0}^{\infty }\hat{\pi}_{n}\left( \rho \right) \tau _{R}^{n}.
\end{equation}%
The calculations become simpler this way (even though less general) as they
allow us to determine the large order order behavior of the gradient series.
In order to avoid confusion with the variables from the previous section,
here we change the notation of the expansion coefficients by adding a hat,
i.e., $\hat{\pi}_{n}$. We note that the coordinates in the line element in %
\eqref{gubserlineelement} are dimensionless\footnote{%
Naturally, an energy scale (called $q$ in \cite{Gubser}) is introduced when
going from $dS_{3}\otimes \mathbb{R}$ back to Minkowski. In this paper we
set this scale to unity.} and so are all the variables computed in $%
dS_{3}\otimes \mathbb{R}$. Therefore, an expansion in powers of $\tau _{R}$
is again an expansion in terms of a dimensionless parameter.

Replacing this expansion in the equation of motion for $\pi $ leads to 
\begin{equation}
\sum_{n=0}^{\infty }\tau _{R}^{n+1}\frac{d\hat{\pi}_{n}}{d\rho }+\frac{2}{3}%
\tanh \rho \sum_{n=0}^{\infty }n\tau _{R}^{n+1}\hat{\pi}_{n}+\sum_{n=0}^{%
\infty }\hat{\pi}_{n}\tau _{R}^{n}+\frac{1}{3}\tanh \rho \sum_{n=0}^{\infty
}\sum_{m=0}^{\infty }\left( 4-n\right) \hat{\pi}_{n}\hat{\pi}_{m}\tau
_{R}^{n+m+1}=\frac{4}{15}\tau _{R}\tanh \rho .
\end{equation}%
We see that the terms can be grouped together according to their power of $%
\tau _{R}$. If we collect all the terms of the same order, we obtain the
equations satisfied by each expansion coefficient $\hat{\pi}_{n}$. We also
note that this procedure of collecting powers of $\tau _{R}$ completely
removes the temperature from the perturbation theory problem. As expected,
the zeroth and first order coefficients satisfy%
\begin{eqnarray}
\hat{\pi}_{0} &=&0, \\
\hat{\pi}_{1}(\rho ) &=&\frac{4}{15}\tanh \rho .
\end{eqnarray}%
The higher-order coefficients obey the equations 
\begin{equation}
\hat{\pi}_{n+1}+\frac{d\hat{\pi}_{n}}{d\rho }+\frac{2n}{3}\tanh \rho \hat{\pi%
}_{n}+\frac{1}{3}\tanh \rho \sum_{m=0}^{n}\left( 4-n+m\right) \hat{\pi}_{n-m}%
\hat{\pi}_{m}=0.
\end{equation}%
More specifically, the second-order coefficient is given by%
\begin{equation}
\hat{\pi}_{2}(\rho )=-\frac{d\hat{\pi}_{1}}{d\rho }-\frac{2}{3}\hat{\pi}%
_{1}\tanh \rho =-\frac{4}{15}+\frac{4}{45}\tanh ^{2}\rho .
\end{equation}%
We note that the solutions obtained for $\hat{\pi}_{1}$ and $\hat{\pi}_{2}$
are exactly the same as the ones obtained in the previous section, provided
one multiplies each coefficient by the appropriate power of the relaxation
time and sets $\epsilon =1$ in the results obtained in the previous section.
However, the procedure described in this section is much simpler to
implement than the one constructed in the previous section. Nevertheless, it
is important to remember that the formalism constructed before is more
general since it is not always possible to re-arrange the problem in terms
of a simpler (though equivalent) perturbative expansion. When dealing with
more general flow configurations or with more complicated perturbative
series (see Sec.\ \ref{sec:slowroll}), one must follow the general procedure
outlined in the previous section.

When $\rho \rightarrow \infty $ all derivatives of $\tanh \rho $ vanish and
the recurrence relations satisfied by $\hat{\pi}_{n}$ considerably
simplifies 
\begin{equation}
\hat{\pi}_{n+1}+\frac{2n}{3}\hat{\pi}_{n}+\frac{1}{3}\sum_{m=0}^{n}\left(
4-n+m\right) \hat{\pi}_{n-m}\hat{\pi}_{m}=0.
\end{equation}%
In this case, the coefficients $\hat{\pi}_{n}$ are just pure numbers that
satisfy recurrence relations that are very similar to those obtained in the
Bjorken case (see, e.g. \cite{Denicol:2017lxn}). In this case, it is
straightforward to see that when $n\gg 1$ this expression leads to $\hat{\pi}%
_{n}\sim n!$ and, consequently, to a divergent series.

\subsection{Divergence of the gradient expansion in Gubser flow}

\begin{figure}[th]
\includegraphics[width=0.5\textwidth]{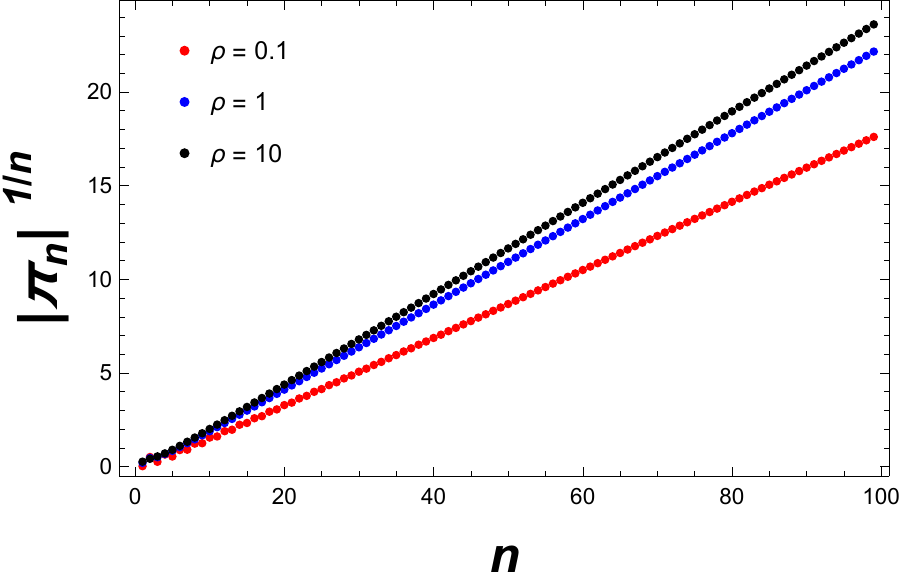}
\caption{(Color online) $|\protect\pi_{n}|^{1/n}$ as a function of $n$ for $%
\protect\rho=0.1$ (red), $\protect\rho=1$ (bule), and $\protect\rho=10$
(black).}
\label{Div1}
\end{figure}

\begin{figure}[th]
\includegraphics[width=0.5\textwidth]{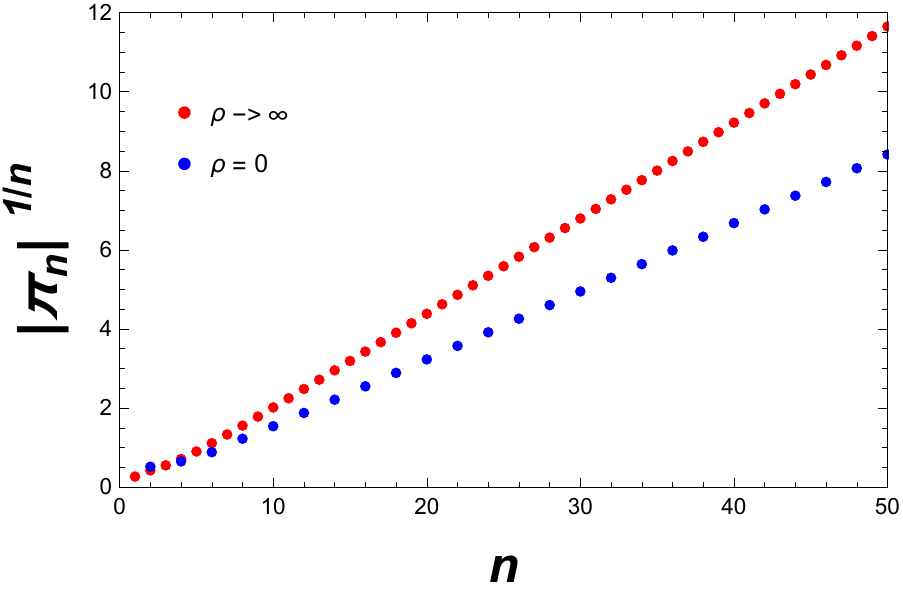}
\caption{(Color online) $|\protect\pi _{n}|^{1/n}$ as a function of $n$ for $%
\protect\rho =0$ and $\protect\rho \to \infty$.}
\label{Div2}
\end{figure}

With the expansion constructed above we are able to determine the large
order behavior of the gradient expansion for arbitrary values of $\rho $.
The recurrence relations derived above were solved with Wolfram's
Mathematica, up to $n=100$, and plotted in Fig.\ \ref{Div1} for $\rho =0.1$, 
$\rho =1$, and $\rho =10$. In Fig.\ \ref{Div2}, we also show the
corresponding result when $\rho =0$ and $\rho \rightarrow \infty $. We do
not plot results for negative values of $\rho $ since the modulus of each
coefficient does not depend on the sign of $\rho $. Both figures clearly
display the factorial growth of the coefficients of the series, indicating
that the gradient expansion has zero radius of convergence. We also remark
that the results vary very little when $\rho $ is changed.

It is interesting to see that when $\rho =0$ (i.e., when the shear tensor is
zero) the gradient expansion still diverges. Moreover, since the first order
term in the expansion is proportional to $\tanh \rho $, this term vanishes
when $\rho =0$ and, as a matter of fact, one can show that all odd terms in
the series vanish in this case. However, the coefficients $\hat{\pi}_{2n}(0)$
do not vanish and these terms alone display factorial growth.

\subsection{Determining the domain of applicability of the gradient expansion%
}

\begin{figure*}[tbp]
\includegraphics[width=0.5\textwidth]{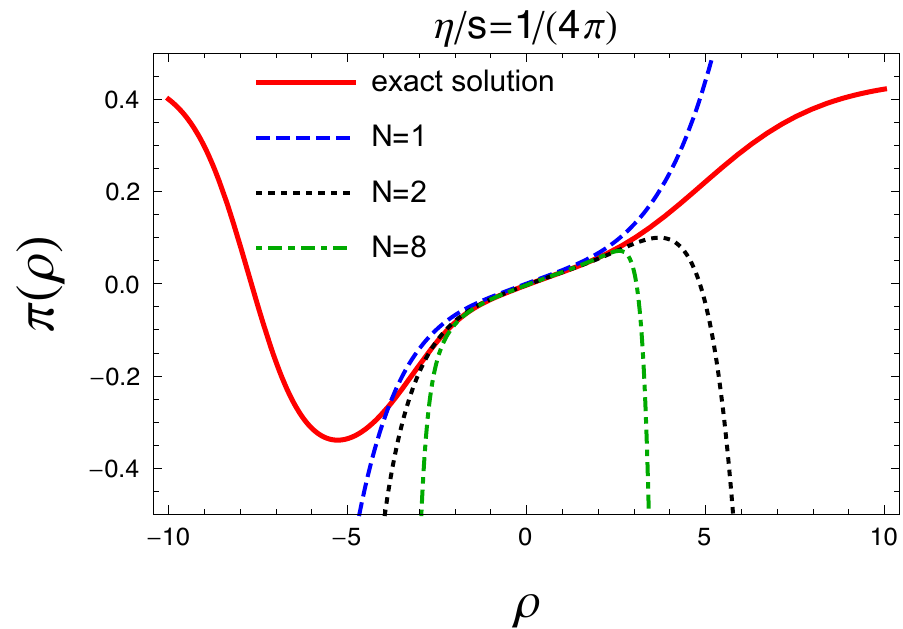}
\caption{(Color online) Comparison between the exact result for $%
\protect\pi$ defined in \eqref{definepinew} obtained by solving Eqs.\ 
\eqref{Good} and \eqref{Great} and the gradient expansion truncated at
different orders. In this plot $\protect\eta/s=1/(4\protect\pi)$. }
\label{Comp0}
\end{figure*}

\begin{figure*}[tbp]
\includegraphics[width=0.5\textwidth]{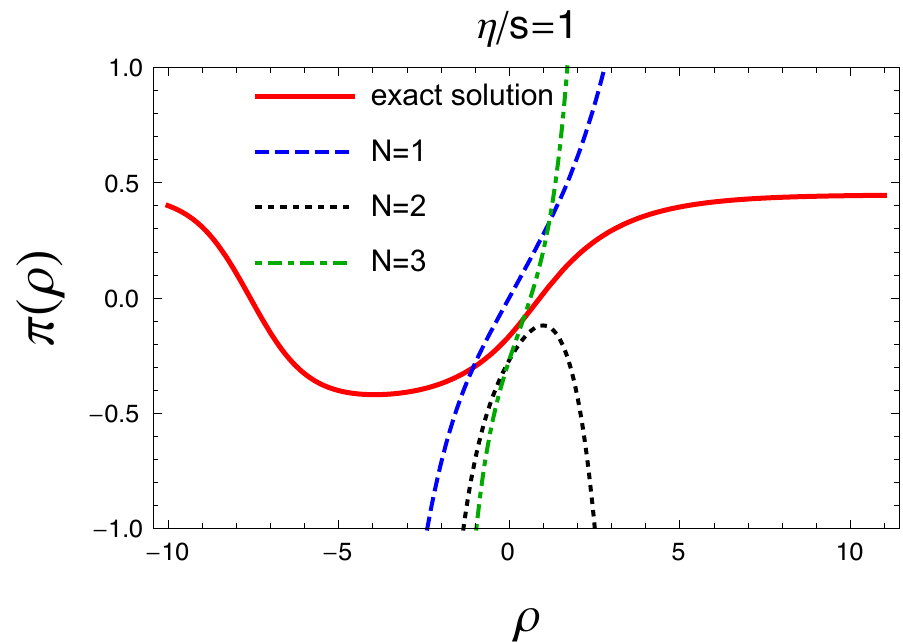}
\caption{(Color online) Comparison between the exact result for $%
\protect\pi$ defined in \eqref{definepinew} obtained by solving Eqs.\ 
\eqref{Good} and \eqref{Great} and the gradient expansion truncated at
different orders. In this plot $\protect\eta/s=1$.}
\label{Comp1}
\end{figure*}

As already mentioned, low order truncations of a divergent expansion can
still be used to provide reasonable approximations for solutions of the
theory, at least in some domains. In Figs.\ \ref{Comp0} and \ref{Comp1} we
compare several truncations of this divergent series with exact solutions
obtained by numerically solving the Israel-Stewart equations \eqref{Good}
and \eqref{Great}, for two values of $\eta /s$ -- $\eta /s=1/\left( 4\pi
\right) $ and $1$. The initial conditions for the numerical problem where
chosen to be $T(\rho _{0})=0.0057$ and $\pi (\rho _{0})=0.4$, with $\rho
_{0}=-30$. We remark that our results do not depend strongly on the value
chosen for $\pi (\rho _{0})$, but they do depend on the choice of $T(\rho
_{0})$ \footnote{%
It is straightforward to see from the equations of motion for $T$ and $\pi $
that rescaling the initial value of the temperature is equivalent to
changing the shear viscosity of the system. This happens in such a way that
reducing the initial temperature of the system corresponds to effectively
increasing the value of $\eta /s$.}. In this comparison, we take the exact
temperature profile $T(\rho )$ and insert it into the corresponding
constitutive relations obtained for $\pi $ using the gradient expansion.

We find that, when the viscosity is small ($\eta /s=1/4\pi $), the 1st and
2nd order truncations of the gradient expansion provide a good approximation
to the exact solution in a wide region around $\rho =0$. We remark that the
best approximation to the exact solution, for this value of viscosity, is
obtained by truncating the expansion at second order, as also happened when
performing this analysis assuming Bjorken flow \cite{Denicol:2017lxn}.
However, the divergent nature of the series is manifest by the fact that the
8th order truncation is significantly worse than the lower orders,
indicating that the optimal truncation of the series is indeed at a lower
order.

Meanwhile, for a larger value of viscosity, $\eta /s=1$, none of the
different truncations of the gradient series is able to provide a reasonable
description of the exact solution. In fact, in this case one even finds that 
$\pi (0)$ deviates significantly from zero in the exact solution -- a result
that is very hard to describe using the gradient expansion, since it implies
that the NS limit is not approached at all even when $\rho \approx 0$. In
general, our results suggests that truncations of the gradient expansion can
provide a good description of solutions of Israel-Stewart theory around $%
\rho =0$, though how far in $\rho $ this occurs (or if this occurs at all)
clearly depends on the value of $\eta /s$.

Furthermore, we remark that the gradient expansion cannot describe the
quantitative and qualitative behavior of the solution when $|\rho |$ is very
large. In fact, it is known from \cite{Marrochio:2013wla} that the exact
solution for $\pi $ in Israel-Stewart theory asymptotes to a constant when $%
|\rho |\gg 1$ -- a result that can never be obtained within a gradient
expansion. However, this is not the limit where one would expect the
gradient expansion to be useful since in this case all powers of $\tanh \rho 
$ become of the same order. In fact, in this regime a complementary
perturbative expansion must be developed, which will be the subject of the
next section.

\section{Expansion in Powers of the Inverse Relaxation Time}

\label{sec:coldplasma}

Previously, we showed that the gradient expansion in Israel-Stewart theory
undergoing Gubser flow corresponds to a series in powers of the relaxation
time. The large order behavior of this expansion, investigated for the first
time in the last section, suggests that it has a zero radius of convergence.
From the nature of the equations, it is obvious that if a series in powers
of $\tau _{R}$ is possible, then a series in powers of $1/\tau _{R}$ should
also be possible, though its regime of applicability should be complementary
to the former. Such a series is interesting in its own way and here we study
the properties of this other type of perturbative expansion. We note that in Section 4.2.1 of Ref.\ \cite{Behtash:2017wqg} a similar series to the one studied in this section was investigated.

Let us now consider again the Israel-Stewart equations and develop an
expansion of $\pi $ in powers of $\tau _{R}^{-1}$%
\begin{equation}
\pi =\sum_{n=0}^{\infty }\tilde{\pi}_{n}\tau _{R}^{-n}.
\end{equation}%
Again, we change our notation for the expansion coefficient to $\tilde{\pi}%
_{n}$ in order to avoid confusion with the variables employed in the
previous sections. Substituting this Ansatz into the simplified
Israel-Stewart equation for $\pi $ \eqref{Great} leads to%
\begin{eqnarray}
&&\sum_{n=0}^{\infty }\tau _{R}^{-n}\frac{d\tilde{\pi}_{n}}{d\rho }%
+\sum_{n=0}^{\infty }\tau _{R}^{-\left( n+1\right) }\tilde{\pi}_{n}-\frac{2}{%
3}\tanh \rho \sum_{n=0}^{\infty }\tau _{R}^{-n}n\tilde{\pi}_{n}  \notag \\
&&+\frac{1}{3}\tanh \rho \sum_{n=0}^{\infty }\sum_{m=0}^{\infty }\tau
_{R}^{-\left( n+m\right) }\left( n+4\right) \tilde{\pi}_{n}\tilde{\pi}_{m}-%
\frac{4}{15}\tanh \rho  \notag \\
&=&0.
\end{eqnarray}

We now follow the same procedure as before and collect the terms that are of
the same order in the inverse relaxation time. At zeroth order we find the
following differential equation of motion for $\tilde{\pi}_{0}$%
\begin{equation*}
\frac{d\tilde{\pi}_{0}}{d\rho }+\frac{4}{3}\tilde{\pi}_{0}^{2}\tanh \rho =%
\frac{4}{15}\tanh \rho .
\end{equation*}%
This equation corresponds to the cold plasma limit solution first found and
studied in Ref.\ \cite{Marrochio:2013wla}, whose analytical solution is 
\begin{equation*}
\tilde{\pi}_{0}(\rho )=\frac{\sqrt{5}}{5}\tanh \left[ \frac{\sqrt{5}}{5}%
\left( \frac{4}{3}\ln \cosh \rho -5b\right) \right] ,
\end{equation*}%
where $b$ is a free parameter that is fixed by the initial condition chosen
for $\pi $. Approximating $\pi $ by this zeroth order solution, one then
obtains the following solution for the temperature,%
\begin{equation*}
T_{0}(\rho )=a\frac{\exp \left( 5b/2\right) }{\left( \cosh \rho \right)
^{2/3}}\cosh ^{1/4}\left[ \frac{\sqrt{5}}{5}\left( \frac{4}{3}\ln \cosh \rho
-5b\right) \right] ,
\end{equation*}%
where $a$ is a free parameter that fixes the initial temperature.

The equation satisfied by the first order coefficient is%
\begin{equation}
\frac{d\tilde{\pi}_{1}}{d\rho }+\tilde{\pi}_{0}+\left( 3\tilde{\pi}_{0}-%
\frac{2}{3}\right) \tilde{\pi}_{1}\tanh \rho =0.
\end{equation}%
This equation does not have a simple analytical solution but it can be
easily solved numerically. Finally, the equation of motion for $\tilde{\pi}%
_{n}$, $n\geq 1$, is%
\begin{equation*}
\frac{d\tilde{\pi}_{n}}{d\rho }+\tilde{\pi}_{n-1}-\left( \frac{2n}{3}\tanh
\rho \right) \tilde{\pi}_{n}+\tanh \rho \sum_{m=0}^{n}\frac{n-m+4}{3}\tilde{%
\pi}_{n-m}\tilde{\pi}_{m}=0.
\end{equation*}%
One can see that this type of expansion is qualitatively different than the
gradient expansion as it requires solving a differential equation at every
order, which makes it harder to determine its large order behavior. In fact,
this introduces back into the problem the initial condition for $\pi $,
which now can be taken into account by the 0th order term of the expansion.
Moreover, differently than the gradient expansion, we note that in this case
there is no approximate constitutive relation between the shear stress
tensor and the hydrodynamic variables.

\begin{figure}[th]
\includegraphics[width=0.5\textwidth]{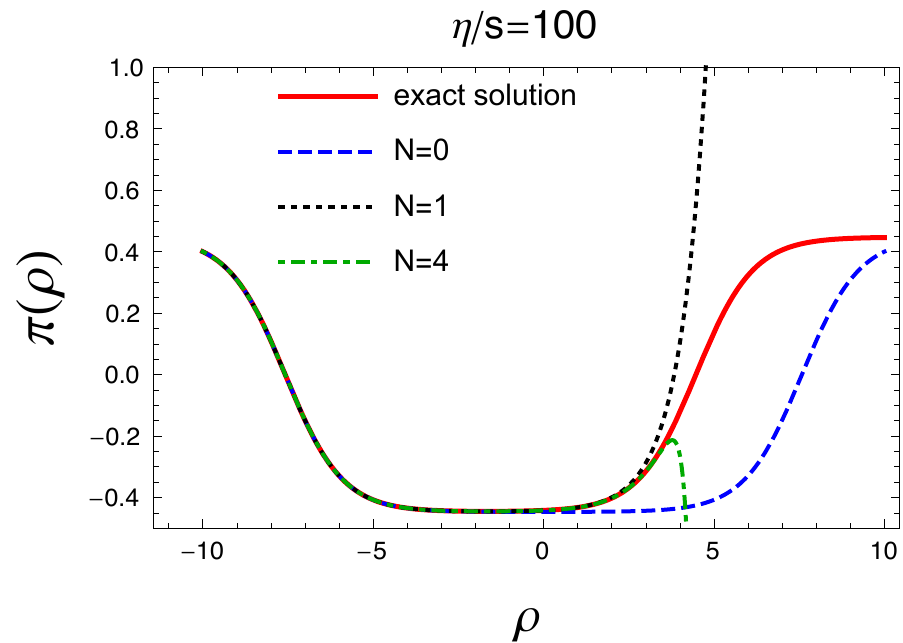}
\caption{(Color online) Comparison between the exact result for $%
\protect\pi$ defined in \eqref{definepinew} obtained by solving Eqs.\ 
\eqref{Good} and \eqref{Great} and the series in inverse powers of the
relaxation truncated at different orders. In this plot $\protect\eta/s=100$.}
\label{Comp3}
\end{figure}

As this series is defined by powers of the inverse relaxation time, if one
keeps the initial condition for the temperature fixed, the regime of
applicability of the series is controlled by how large $\eta /s$ is. In
Fig.\ \ref{Comp3} we set $\eta /s=100$ and compare the exact solution of the
Israel-Stewart equations with the same initial conditions as before to
different truncations of this new series in powers of the inverse relaxation
time. One can see that the 0th order term already correctly describes the
asymptotic regime at very large $|\rho |$, though the agreement worsens for $%
0<\rho <10$. The inclusion of higher order terms improves the agreement near 
$\rho \sim 0$ though for $5<\rho <10$ not even the 4th order truncation can
describe the numerical solution.

We finish this section with the remark that in flow situations with less
symmetry, the type of series developed in this section basically requires
solving a problem as hard as the original problem (coupled, nonlinear
partial differential equations for the shear stress tensor) at each order.
Therefore, we expect its use in practical applications to be more limited
than the more easily implementable gradient expansion. In the next section
we develop a perturbative expansion that can describe not only the $\rho\sim
0$ regime but also the asymptotic values of the solution when $\rho \to
\infty$.

\section{Slow-Roll Expansion}

\label{sec:slowroll}

The slow-roll expansion was first used in the context of hydrodynamics in 
\cite{Heller:2015dha} in the case of Bjorken flow. The systematic
implementation of this series in that problem was discussed in \cite%
{Strickland:2017kux} and later in \cite{Denicol:2017lxn}. The slow-roll
expansion in Gubser flow is defined by the following perturbative problem 
\begin{equation}
\epsilon \,c\frac{d\pi }{d\rho }+T\pi +\frac{4}{3}c\pi ^{2}\tanh \rho =\frac{%
4}{15}c\tanh \rho ,  \label{LULALIVRE}
\end{equation}%
where now the book-keeping parameter $\epsilon $ multiplies only the
derivative term of the equation. As before, we look for perturbative
solutions of the form 
\begin{equation}
\pi \sim \sum_{n=0}^{\infty }\bar{\pi}_{n}\epsilon ^{n},
\end{equation}%
and, at the end of the calculation, set $\epsilon =1$, recovering, in
principle, a solution of the original equation of motion. Replacing this
expansion into Eq.~\eqref{LULALIVRE}, we obtain the following set of
relations 
\begin{equation}
c\sum_{n=0}^{\infty }\frac{d\bar{\pi}_{n}}{d\rho }\epsilon
^{n+1}+T\sum_{n=0}^{\infty }\bar{\pi}_{n}\epsilon ^{n}+\frac{4}{3}%
c\sum_{n=0}^{\infty }\sum_{m=0}^{\infty }\bar{\pi}_{n}\bar{\pi}_{m}\epsilon
^{n+m}\tanh \rho =\frac{4}{15}c\tanh \rho .
\end{equation}%
As before, we write the derivative of $\bar{\pi}_{n}$ as 
\begin{equation*}
\frac{d\bar{\pi}_{n}}{d\rho }=\left. \frac{\partial \bar{\pi}_{n}}{\partial
\rho }\right\vert _{T}+\left. \frac{\partial \bar{\pi}_{n}}{\partial T}%
\right\vert _{\tanh \rho }\frac{dT}{d\rho },
\end{equation*}%
which allows us to properly collect all powers of $\epsilon $ and leads to
the following equations, 
\begin{gather}
c\sum_{n=0}^{\infty }\left( \left. \frac{\partial \bar{\pi}_{n}}{\partial
\rho }\right\vert _{T}-\frac{2}{3}\tanh \rho \,\left. T\frac{\partial \bar{%
\pi}_{n}}{\partial T}\right\vert _{\tanh \rho }\right) \epsilon ^{n+1}+\frac{%
c}{3}\tanh \rho \sum_{n=0}^{\infty }\sum_{m=0}^{\infty }\bar{\pi}_{m}T\left. 
\frac{\partial \bar{\pi}_{n}}{\partial T}\right\vert _{\tanh \rho }\epsilon
^{n+m+1}  \notag \\
+T\sum_{n=0}^{\infty }\bar{\pi}_{n}\epsilon ^{n}+\frac{4}{3}%
c\sum_{n=0}^{\infty }\sum_{m=0}^{\infty }\bar{\pi}_{n}\bar{\pi}_{m}\epsilon
^{n+m}\tanh \rho =\frac{4}{15}c\tanh \rho .
\end{gather}

The zeroth order solution is the Gubser flow generalization of the
well-known result first derived for Bjorken flow in \cite{Heller:2015dha}, 
\begin{gather}
\frac{4}{3}\bar{\pi}_{0}^{2}\tau _{R}\tanh \rho +\bar{\pi}_{0}=\frac{4}{15}%
\tau _{R}\tanh \rho ,  \notag \\
\Longrightarrow \bar{\pi}_{0}^{\pm }(\rho )=\frac{-3\pm \sqrt{9+\frac{4}{5}%
\left( 4\tau _{R}\tanh \rho \right) ^{2}}}{8\tau _{R}\tanh \rho }.
\label{definepi0slowroll}
\end{gather}%
We remark that the 0th order term of the slow-roll expansion for
Israel-Stewart theory shown above was first computed in \cite%
{Behtash:2017wqg}. Out of the two possible solutions obtained above, we
chose the one which asymptotes to the Navier-Stokes solution in the limit $%
\rho \rightarrow 0$, i.e., we consider only the solution $\bar{\pi}_{0}^{+}$%
. In fact, one may expand \eqref{definepi0slowroll} in powers of $\tau _{R}$
to find 
\begin{equation}
\bar{\pi}_{0}^{+}(\rho )=\frac{4}{15}\tau _{R}\tanh \rho -\frac{64}{675}%
\left( \tau _{R}\tanh \rho \right) ^{3}+\mathcal{O}(\tau _{R}^{5}).
\end{equation}%
Thus, we see that the 0th order term in the slow-roll expansion recovers the
NS result (i.e., it matches the gradient expansion truncated at 1st order).
However, it is important to notice that the higher order terms generated by
Taylor expanding $\bar{\pi}_{0}^{+}$ differ from the higher order terms
present in the gradient expansion.

Also, it is interesting to notice that the 0th order slow-roll term $\bar{\pi%
}_{0}^{+}$ is solely a function of the Knudsen number $\tau _{R}\tanh \rho \sim
\tau _{R}\sqrt{\sigma _{\mu \nu }\sigma ^{\mu \nu }}$ for Gubser flow, as
also happened with the first order truncation of the gradient expansion (NS
theory). A similar situation was found in Bjorken flow, where the
zeroth--order truncation of the slow--roll expansion was shown to depend
solely on the combination $T\tau $ -- the inverse Knudsen number for this
flow configuration \cite{Heller:2015dha}. In Bjorken flow this feature
persisted to all orders in the slow-roll expansion \cite{Denicol:2017lxn}
though here we shall see that this does not hold for the slow-roll expansion
in Gubser flow (the same was observed for the gradient expansion in Gubser
flow in Sec.\ \ref{gradientscheme1}). As a matter of fact, we shall
explicitly show in the following that the higher order terms of the
slow-roll series in Gubser flow are functions of both $\tau _{R}$ and $\tanh
\rho $, separately.

Using Eq.\ \eqref{definepi0slowroll}, higher order solutions are obtained by
solving the recurrence relation%
\begin{gather}
\left. \frac{\partial \bar{\pi}_{n}}{\partial \rho }\right\vert _{T}-\frac{2T%
}{3}\tanh \rho \left. \frac{\partial \bar{\pi}_{n}}{\partial T}\right\vert
_{\tanh \rho }+\frac{T}{3}\tanh \rho \sum_{m=0}^{n}\bar{\pi}_{m}\left. \frac{%
\partial \bar{\pi}_{n-m}}{\partial T}\right\vert _{\tanh \rho }  \notag \\
+\frac{T}{c}\bar{\pi}_{n+1}+\frac{4}{3}\sum_{m=0}^{n+1}\bar{\pi}_{n+1-m}\bar{%
\pi}_{m}\tanh \rho =0,  \label{FORATEMER}
\end{gather}%
which can be simplified to%
\begin{eqnarray}
\left( \frac{1}{\tau _{R}}+\frac{8}{3}\bar{\pi}_{0}^{+}\tanh \rho \right) 
\bar{\pi}_{n+1} &=&-\frac{2}{3}\tanh \rho \left. \tau _{R}\frac{\partial 
\bar{\pi}_{n}}{\partial \tau _{R}}\right\vert _{\tanh \rho }-\left. \frac{%
\partial \bar{\pi}_{n}}{\partial \rho }\right\vert _{\tau _{R}}+\frac{1}{3}%
\tanh \rho \sum_{m=0}^{n}\bar{\pi}_{m}\left. \tau _{R}\frac{\partial \bar{\pi%
}_{n-m}}{\partial \tau _{R}}\right\vert _{\tanh \rho }  \notag \\
&&-\frac{4}{3}\sum_{m=1}^{n}\bar{\pi}_{n+1-m}\bar{\pi}_{m}\tanh \rho .
\label{definepinSR}
\end{eqnarray}%
Solving this recurrence relation, one can determine the first order solution
to be 
\begin{equation}
\bar{\pi}_{1}(\rho )=\frac{15\bar{\pi}_{0}^{+}}{\tau _{R}\tanh \rho }\left[ 
\frac{\left( 1+\bar{\pi}_{0}^{+}\right) \left( \tau _{R}\tanh \rho \right)
^{2}-3\tau _{R}^{2}}{45+64\left( \tau _{R}\tanh \rho \right) ^{2}}\right] .
\label{MOROBANDIDO}
\end{equation}%
This result clearly demonstrates that $\bar{\pi}_{1}$ depends on both $\tau
_{R}$ and $\tanh \rho $, separately, a fact that shall remain true for all
higher order coefficients of the slow-roll expansion.

Furthermore, expanding now the truncated series at 1st order (i.e., $\bar{\pi%
}\rightarrow \bar{\pi}_{0}^{+}+\bar{\pi}_{1}$) in powers of $\tau _{R}$
leads to 
\begin{equation}
\bar{\pi}(\rho )=\frac{4}{15}\tau _{R}\tanh \rho -\frac{4}{15}\left( \tau
_{R}\right) ^{2}\left( 1-\frac{1}{3}\tanh ^{2}\rho \right) -\frac{16}{225}%
(\tau _{R}\tanh \rho )^{3}+\mathcal{O}(\tau _{R}^{4}),
\end{equation}%
which shows that the 1st-order truncation of the slow-roll series is able to
recover the result obtained from the gradient expansion truncated at 2nd
order, see \eqref{grad3rdorder}. In general, a Taylor expansion of the
slow-roll expansion truncated at the $N$--th term will reduce to the correct
expression for the gradient expansion truncated at order $N+1$. In this
sense, the slow-roll expansion can be seen as a type of reorganization of
the gradient series that contains an infinite resummation of gradients at a
given order.

\subsection{Divergence of the slow-roll series in Gubser flow}

\begin{figure}[th]
\includegraphics[width=0.5\textwidth]{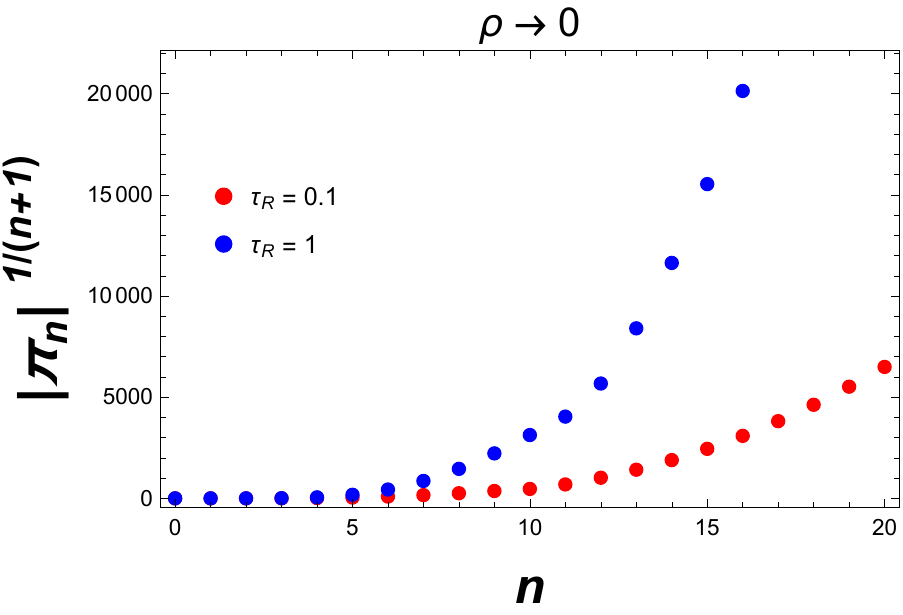}
\caption{(Color online) Large order behavior of the slow-roll series when $%
\protect\rho \rightarrow 0$ and $\protect\tau _{R}=0.1,1$.}
\label{fig:SRrho0vsn}
\end{figure}

\begin{figure}[th]
\includegraphics[width=0.5\textwidth]{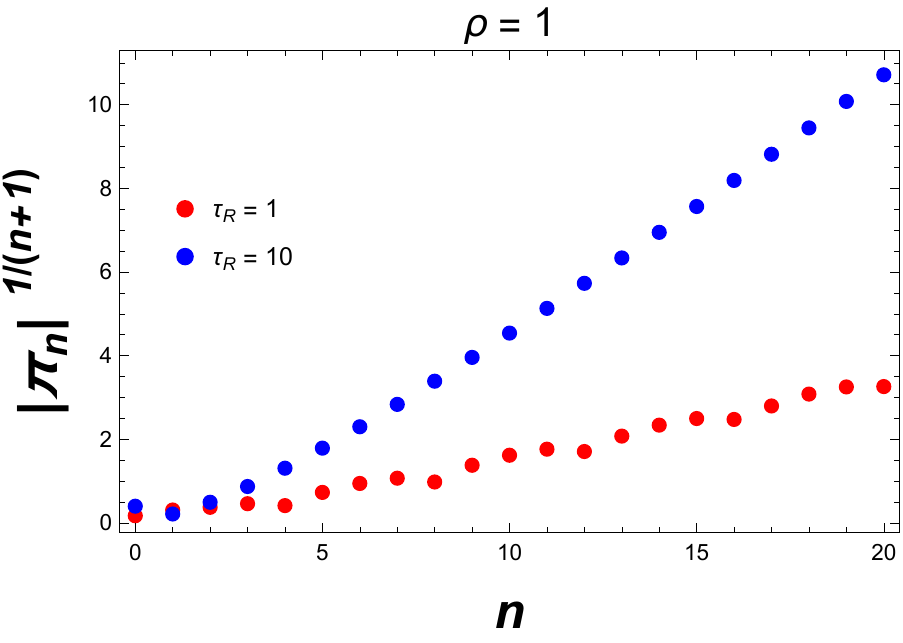}
\caption{(Color online) Large order behavior of the slow-roll series when $%
\protect\rho =1$ and $\protect\tau _{R}=1,10$.}
\label{fig:SRrho1vsn}
\end{figure}

Now we solve Eq.\ \eqref{definepinSR} numerically to determine the behavior
of the slow-roll series at higher orders. In Fig.\ \ref{fig:SRrho0vsn} we
show how the coefficients of the series change with $n$ when $\rho
\rightarrow 0$ and for the following values of relaxation time, $\tau
_{R}=0.1,1$. 
This plot indicates that, in this limit, the magnitude of the terms grows
larger than $n!$ when $n$ is large\footnote{%
We checked that $|\pi _{n}|$ does not grow larger $(n!)^{\alpha }$ where $%
\alpha <1.3$ in the range considered (this result is robust with respect to
the choice of values for $\tau _{R}$).}. To illustrate the fact that the
large order behavior of the slow-roll series now depends on two parameters
(i.e., $\rho $ and $\tau _{R}$), we show in Fig.\ \ref{fig:SRrho1vsn} what
happens when $\rho =1$. Even though the series still appears to diverge, in
this case $|\bar{\pi}_{n}|^{1/(n+1)}$ only grows linearly with $n$ (the same
qualitative result appears for other values of $\tau _{R}$ and also when $%
\rho $ is negative). Therefore, in Israel-Stewart theory both the gradient
and the slow-roll expansions generally diverge in Gubser flow, just as it
occurred in Bjorken flow \cite{Denicol:2017lxn}. 

The only exception occurs when $|\rho |\rightarrow \infty $. Since the
temperature vanishes in this limit one must also take $\tau _{R}\rightarrow
\infty $, which implies that $\bar{\pi}_{0}^{\pm }\rightarrow \pm \mathrm{%
sign}\,\rho /\sqrt{5}$ and Eq.\ \eqref{definepinSR} gives $\bar{\pi}_{n>0}=0$%
. This shows that the slow-roll expansion in fact converges in this limit to 
$\bar{\pi}_{0}^{\pm }$. As discussed in \cite{Marrochio:2013wla}, solutions
of the Israel-Stewart equation for the shear stress tensor \eqref{Great} do
display the same behavior and, thus, one can see that the slow-roll series
necessarily converges to solutions of the Israel-Stewart equation when $%
|\rho |\rightarrow \infty $.

\subsection{Determining the domain of applicability of the slow-roll
expansion}

\begin{figure}[th]
\includegraphics[width=0.5\textwidth]{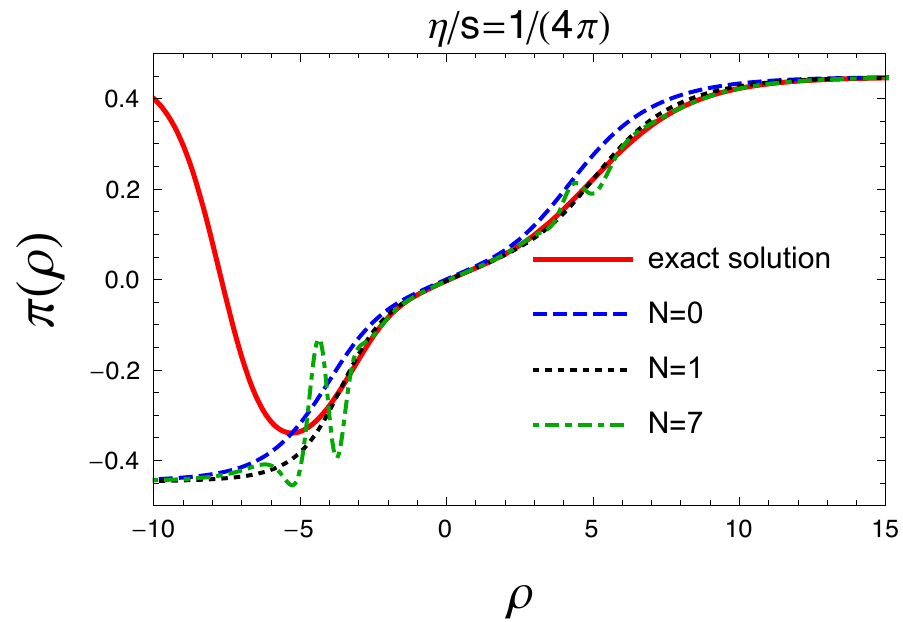}
\caption{(Color online) Comparison between the exact solution of IS
equations and different truncations of the slow-roll series in Gubser flow
for $\protect\eta/s = 1/(4\protect\pi)$.}
\label{fig:compSRetas14pi}
\end{figure}

\begin{figure*}[tbp]
\includegraphics[width=0.5\textwidth]{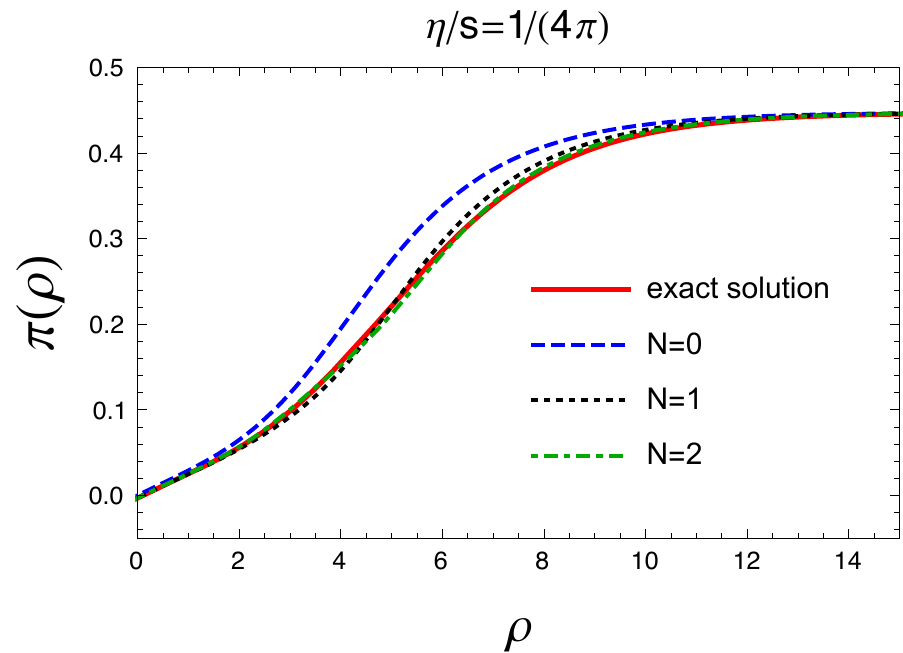}
\caption{(Color online) Detailed comparison when $\protect\rho>0$ between
the exact solution of the IS equations and the low order truncations of
the slow-roll series in Gubser flow for $\protect\eta/s=1/(4\protect\pi)$.}
\label{fig:compSRetas14pizoom}
\end{figure*}

\begin{figure*}[tbp]
\includegraphics[width=0.5\textwidth]{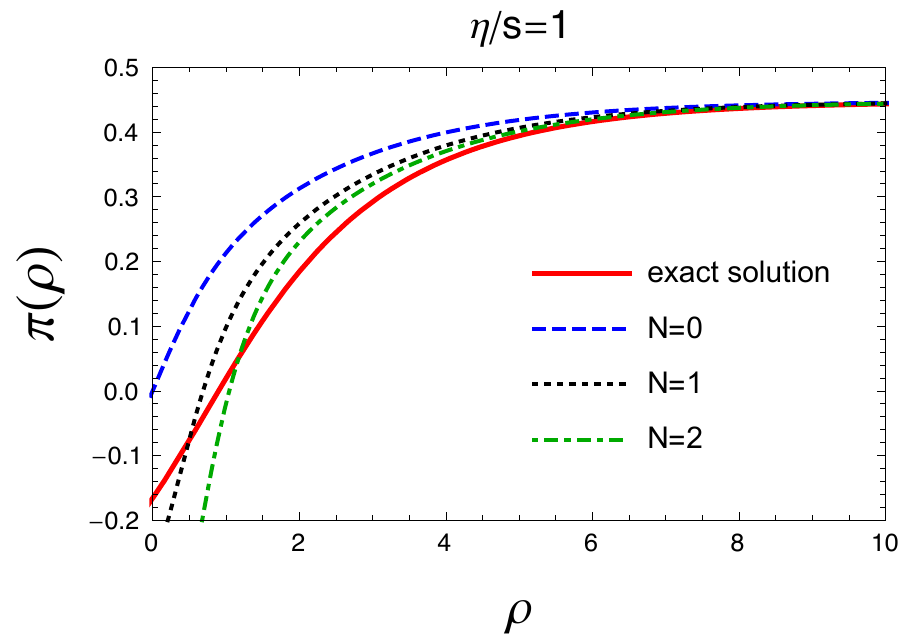}
\caption{(Color online) Detailed comparison when $\protect\rho>0$ between
the exact solution of the IS equations and the low order truncations of
the slow-roll series in Gubser flow for $\protect\eta/s=1$.}
\label{fig:compSRetas1zoom}
\end{figure*}

In Figs.\ \ref{fig:compSRetas14pi}, \ref{fig:compSRetas14pizoom}, \ref%
{fig:compSRetas1}, and \ref{fig:compSRetas1zoom} we investigate how
different truncations of the slow-roll series for $\pi $ fare in comparison
to the same exact solution of the Israel-Stewart equations used in previous
sections. These exact solutions were constructed using the same initial
condition for $T$ and $\pi $ employed in the previous sections. We can see
in Fig.\ \ref{fig:compSRetas14pi} that, for $\eta /s=1/(4\pi )$, the 0th order
truncation deviates only slightly from the exact solution though the 1st order
truncation of the slow-roll expansion gives a reasonably accurate
description of the solution for $\rho >-5$. We also show the results for the
7th order truncation, which are found to display oscillatory behavior
compatible with the divergent character of the series when $\rho $ is finite
(when $|\rho |\rightarrow \infty $, however, the series converges and this
is why the oscillations do not appear in that regime). Figure\ \ref%
{fig:compSRetas14pizoom} shows that the agreement with the exact
solution improves at 2nd order for $\rho >0$.

\begin{figure*}[tbp]
\includegraphics[width=0.5\textwidth]{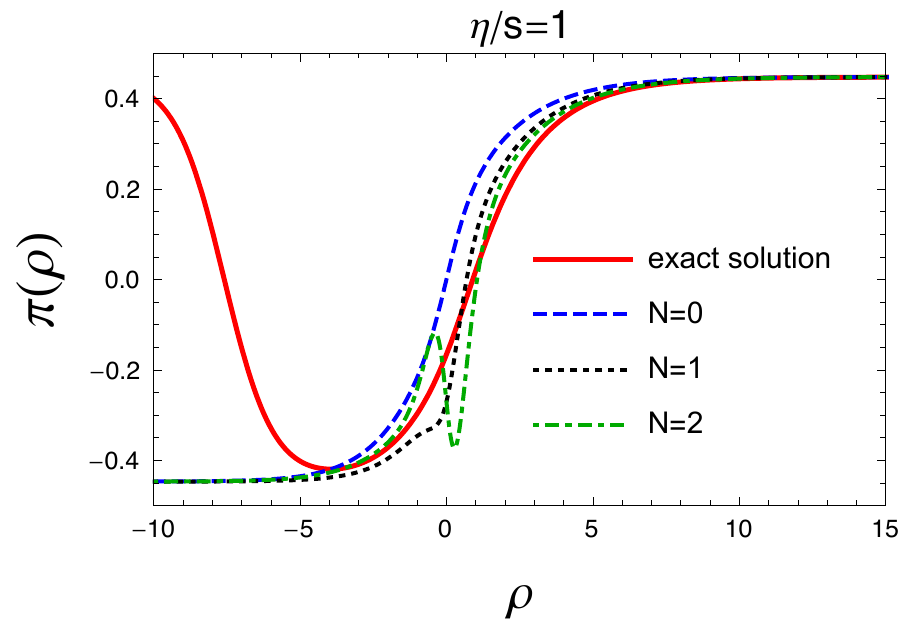}
\caption{(Color online) Comparison between the exact solution of IS
equations and different truncations of the slow-roll series in Gubser flow
for $\protect\eta/s = 1$.}
\label{fig:compSRetas1}
\end{figure*}

Figures\ \ref{fig:compSRetas1} and \ref{fig:compSRetas1zoom} show that the
agreement between the truncated slow-roll series and the exact solution
considerably worsens when $\eta /s=1$ (the same occurred in the gradient
expansion investigated in Sec.\ \ref{gradientscheme2}). As already mentioned, in this
case $\pi (0)$ deviates considerably from zero in the exact solution and the
0th order approximation of the slow-roll series is not accurate even at $%
\rho =0$. This affects the overall ability of the truncated series to
describe the solution and one can see in Fig.\ \ref{fig:compSRetas1} that
oscillations now appear already at 2nd order. Figure\ \ref%
{fig:compSRetas1zoom} shows the results for the $\rho >0$ region in detail.
Even though the higher order truncations become closer to
the exact solution, the agreement is generally poor in comparison to what
was found in Fig.\ \ref{fig:compSRetas14pizoom}, where $\eta /s=1/(4\pi )$.
Overall, since $T$ has a maximum at $\rho =0$ in Israel-Stewart theory \cite%
{Marrochio:2013wla}, the relaxation time has a minimum at the same location
and we find that the slow-roll series is not a good proxy for the exact
solutions when $\tau _{R}(0)\gtrsim 1$, which occurs in this example when $%
\eta /s=1$.

\subsection{Comparison between the perturbative expansions}

\begin{figure}[th]
\includegraphics[width=0.5%
\textwidth]{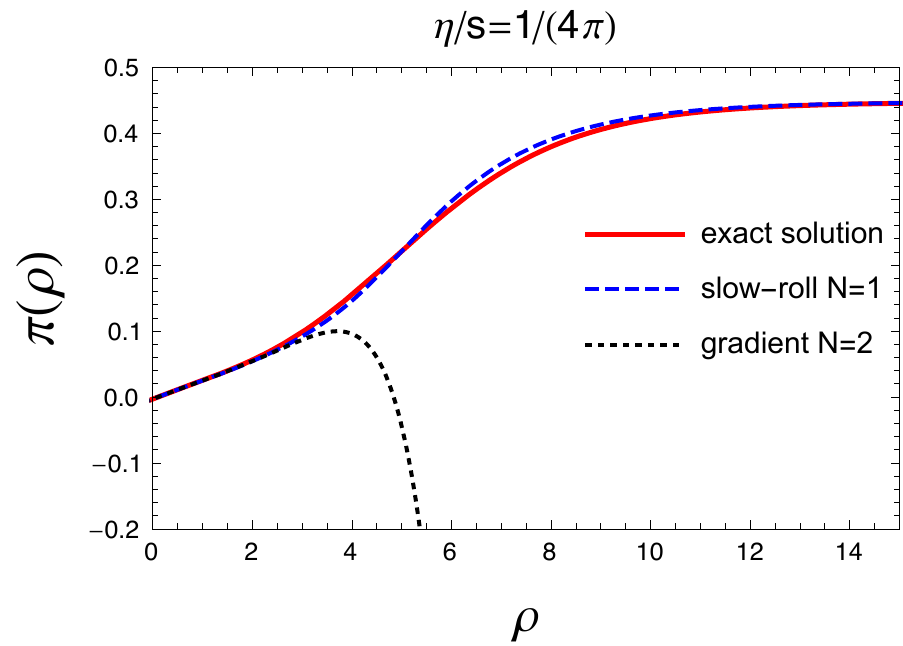}
\caption{(Color online) Detailed comparison when $\protect\rho>0$ between
the exact solution of IS equations undergoing Gubser flow for $\protect%
\eta/s=1/(4\protect\pi)$ and the $N=1$ and $N=2$ truncations of the
slow-roll and gradient series, respectively.}
\label{fig:compgradSRetas14pizoom}
\end{figure}

We mentioned before that a Taylor series in powers of $\tau _{R}$ of the $N$%
-th order truncation of the slow-roll expansion reduces to the result
obtained from the $(N+1)$-th order truncation of the gradient series. In
order to assess how these two perturbative series fare in comparison to the
exact solution of Israel-Stewart equations, we plot in Fig.\ \ref%
{fig:compgradSRetas14pi} the 2nd order gradient expansion result and the $%
N=1 $ truncation of the slow-roll series for $\eta /s=1/(4\pi )$. One can
see that the $N=1$ slow-roll expansion indeed matches the $N=2$ gradient
series result as long as the latter still provides a good approximation to
the exact solution (approximately when $-3<\rho <3$). However, outside this
regime the gradient series fails to describe the solution while the
slow-roll result nicely continues to provide a very good description of the
system's dynamics towards larger values of $\rho $. A detailed comparison
between the truncated series and the numerical result when $\rho >0$ can be
found in Fig.\ \ref{fig:compgradSRetas14pizoom}. The 1st order slow-roll
series is much more accurate in this regime than the gradient expansion,
which fails around $\rho \sim 3$.

\begin{figure}[th]
\includegraphics[width=0.5\textwidth]{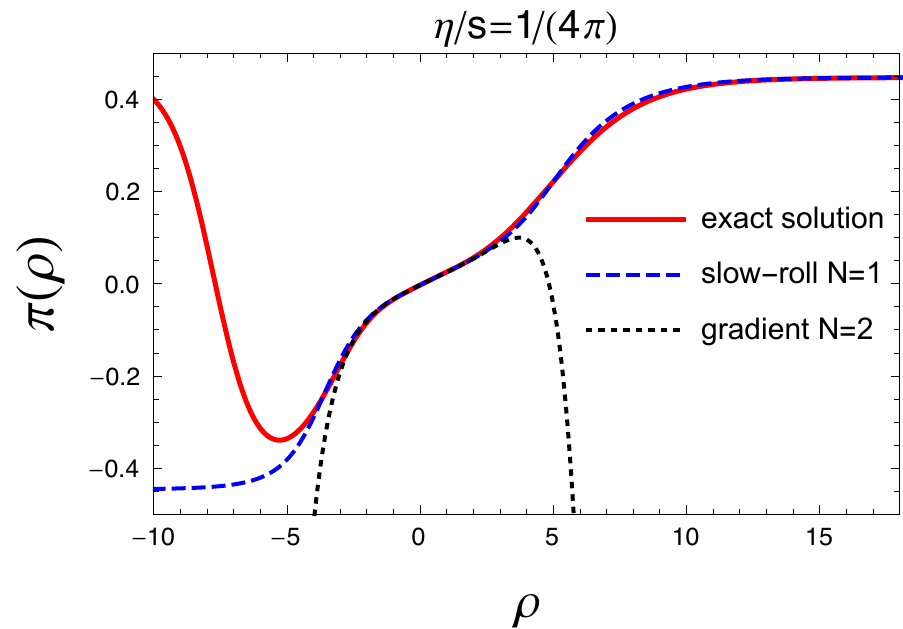}
\caption{(Color online) Comparison between the exact solution of IS
equations undergoing Gubser flow for $\protect\eta/s=1/(4\protect\pi)$ and
the $N=1$ and $N=2$ truncations of the slow-roll and gradient series,
respectively.}
\label{fig:compgradSRetas14pi}
\end{figure}

\begin{figure}[th]
\includegraphics[width=0.5\textwidth]{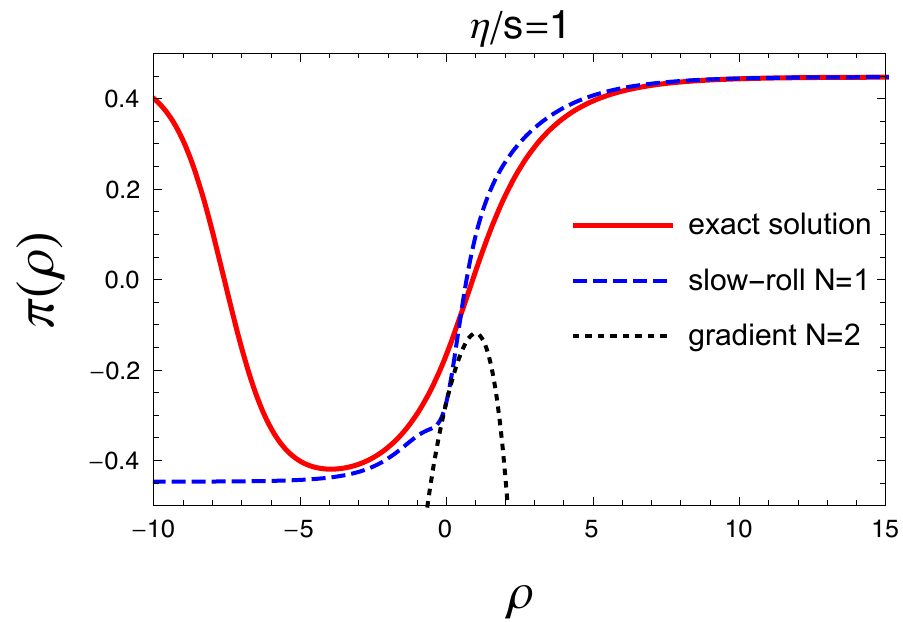}
\caption{(Color online) Comparison between the exact solution of IS
equations undergoing Gubser flow for $\protect\eta/s=1$ and the $N=1$ and $%
N=2$ truncations of the slow-roll and gradient series, respectively.}
\label{fig:compgradSRetas1}
\end{figure}

Figure\ \ref{fig:compgradSRetas1} shows that, when $\eta /s=1$, both
expansions have difficulties in describing the behavior of the exact
solution, as expected from the results of the previous sections. However, we
remark that the 1st order truncation of the slow-roll series still behaves
much better than the 2nd order gradient expansion. In general, perturbative
series such as the gradient or the slow-roll expansions can only be accurate
if their lowest order terms are not too far from the exact solution. This is
the case when $\eta /s=1/(4\pi )$, since $\pi (0)$ is very small and can be
well approximated by the Navier-Stokes solution. The same does not happen
when $\eta /s=1$, in which case the solution for $\pi (0)$ considerably
deviates from zero.

\section{Attractor Solution in Gubser flow}

\label{sec:attractor}

In this section we investigate the attractor solution of Israel-Stewart
equations for a system expanding according to Gubser flow, which was first
studied in Ref.\ \cite{Behtash:2017wqg}. In this work, we shall interpret the attractor 
solution as a resummed slow-roll expansion. The analysis presented in
this section aims to explore two fundamental aspects of the attractor: its possible functional
dependence on the Knudsen number and its approximate description using the slow-roll series
truncated at higher orders.

\begin{figure}[th]
\includegraphics[width=0.5\textwidth]{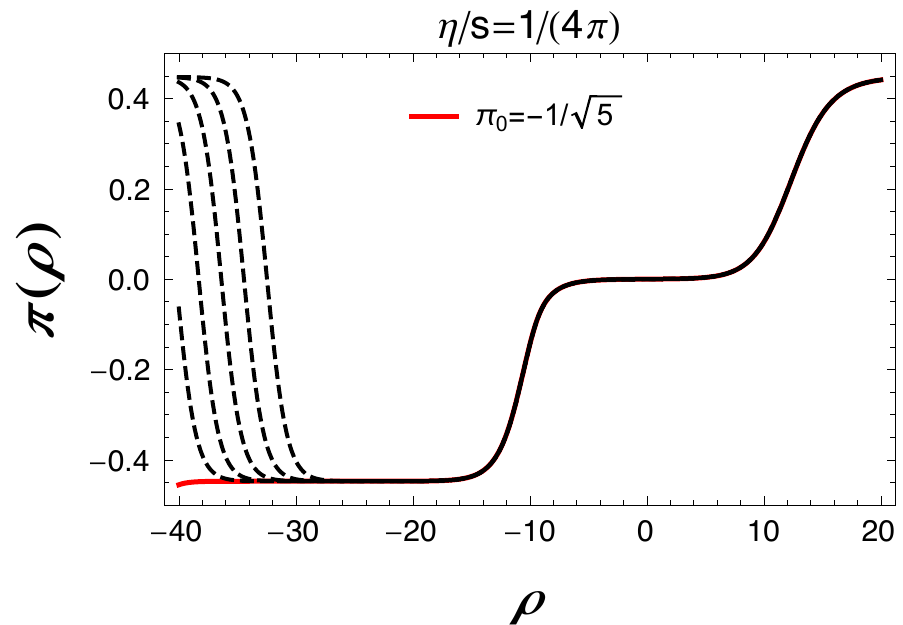}
\caption{(Color online) Attractor solution of IS equations undergoing Gubser
flow (solid red) compared to other solutions of the equations (dashed black
curves) for $\protect\eta/s=1/(4\protect\pi)$.}
\label{fig:pietas14piattractorfixedT}
\end{figure}

\begin{figure}[th]
\includegraphics[width=0.5\textwidth]{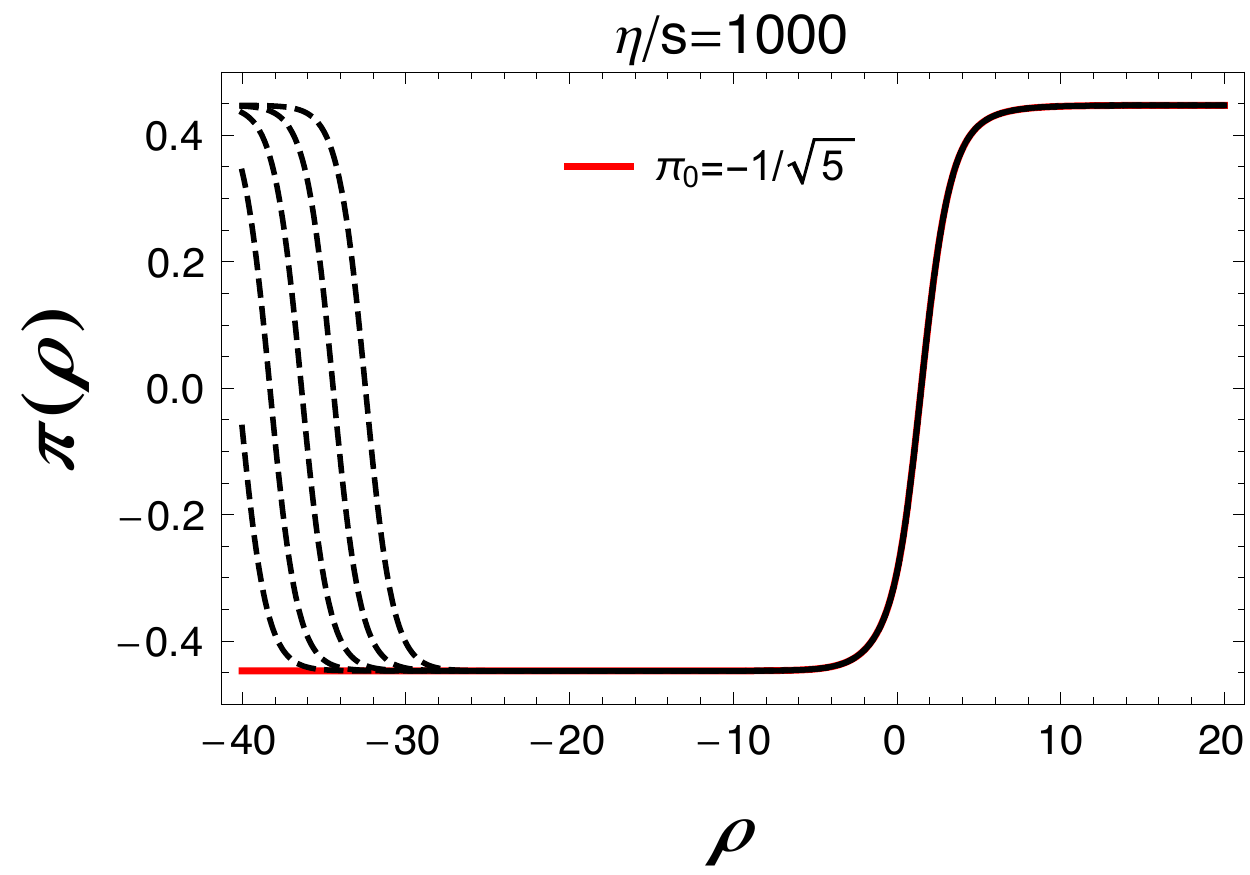}
\caption{(Color online) Attractor solution of IS equations undergoing Gubser
flow (solid red) compared to other solutions of the equations (dashed black
curves) for the very large value of $\protect\eta/s=1000$.}
\label{fig:pietas1000attractorfixedT}
\end{figure}

In the previous section, we found that the slow-roll series converges when $%
|\rho |\rightarrow \infty $ and we can use this fact to numerically
construct a resummed version of the slow-roll solution. We found that there
are two possible convergent solutions for $\pi $ when $|\rho |\rightarrow
\infty $: $1/\sqrt{5}$ and $-1/\sqrt{5}$. It is natural to expect that one
of these boundary conditions will lead to the exact solution related to the
slow-roll expansion. In practice, we observe that the vast majority of
solutions (if not all of them) of Israel-Stewart theory actually converges
to $1/\sqrt{5}$ when $\rho \rightarrow -\infty $ (for $\rho \rightarrow
\infty $, all known numerical solutions of Israel-Stewart theory also
converge to $1/\sqrt{5}$). So far, the only case in which we are able to
obtain a solution that is equal to $-1/\sqrt{5}$, when $\rho \rightarrow
-\infty $, is when we give exactly this boundary condition at a very small
value of $\rho $. Any small deviation from $-1/\sqrt{5}$ will make the
solution tend to $1/\sqrt{5}$ when we decrease the value of $\rho $. This
behavior suggests that the boundary condition $\pi \left( -\infty \right)
=-1/\sqrt{5}$ defines a unique solution of the equation and that such unique
solution can be identified as the resummed result for the slow-roll
series. On the other hand, the other boundary condition at $\rho =-\infty $%
, $\pi \left( -\infty \right) =1/\sqrt{5}$, is satisfied by an infinite
number of solutions and cannot be used to define any specific solutions of
the equations.

This is illustrated in Figs.\ \ref{fig:pietas14piattractorfixedT} and \ref%
{fig:pietas1000attractorfixedT} for two vastly different values of $\eta /s$%
. In these plots, the solid red curve depicts the solution of Israel-Stewart
equations assuming $T(-30)=9.222\times 10^{-8}$ and $\pi (-30)=-1/\sqrt{5}$.
The dashed curves are computed keeping the initial condition for the
temperature fixed while considering very small variations of $\pi (-30)$, of
at most 1\% around $-1/\sqrt{5}$. We see that any small variation of the
value of $\pi $ at $\rho =-30$ makes the solution converge to $1/\sqrt{5}$
when $\rho $ is decreased. This only does not happen when we fix $\pi $ to
be exactly $-1/\sqrt{5}$ (red curve). Furthermore, one can see that all the
solutions converge extremely rapidly to the solution where $\lim_{\rho
\rightarrow -\infty }\pi (\rho )=-1/\sqrt{5}$. This solution, represented here by
the solid red curve in these plots, corresponds to the hydrodynamic attractor solution
first discussed in \cite{Heller:2015dha} in Bjorken flow, which was later
investigated in more detail in \cite{Strickland:2017kux}.

Comparing Figs.\ \ref{fig:pietas14piattractorfixedT} and \ref%
{fig:pietas1000attractorfixedT} we see that the profile of the attractor
depends on the value of $\eta /s$, becoming closer to a step function as $%
\eta /s$ is increased even further. Such a behavior is very hard to describe
using the slow-roll series, as one can see in Fig.\ \ref%
{fig:pietascomparaattractor}. In this figure we compare the attractor
solutions (solid red) with truncations of the slow-roll series. For $\eta
/s=1/(4\pi )$ we used the 2nd order truncation while for the extremely large
value of $\eta /s=1000$ we only took the 0th order term in the series. When $%
\eta /s$ is small, the truncated slow-roll series provides an excellent
description of the attractor while for very large values of $\eta /s$ this
perturbative approach provides a poor description, even though still qualitatively
accurate, of the attractor away from the asymptotic regime, as expected.

\begin{figure}[th]
\includegraphics[width=0.5\textwidth]{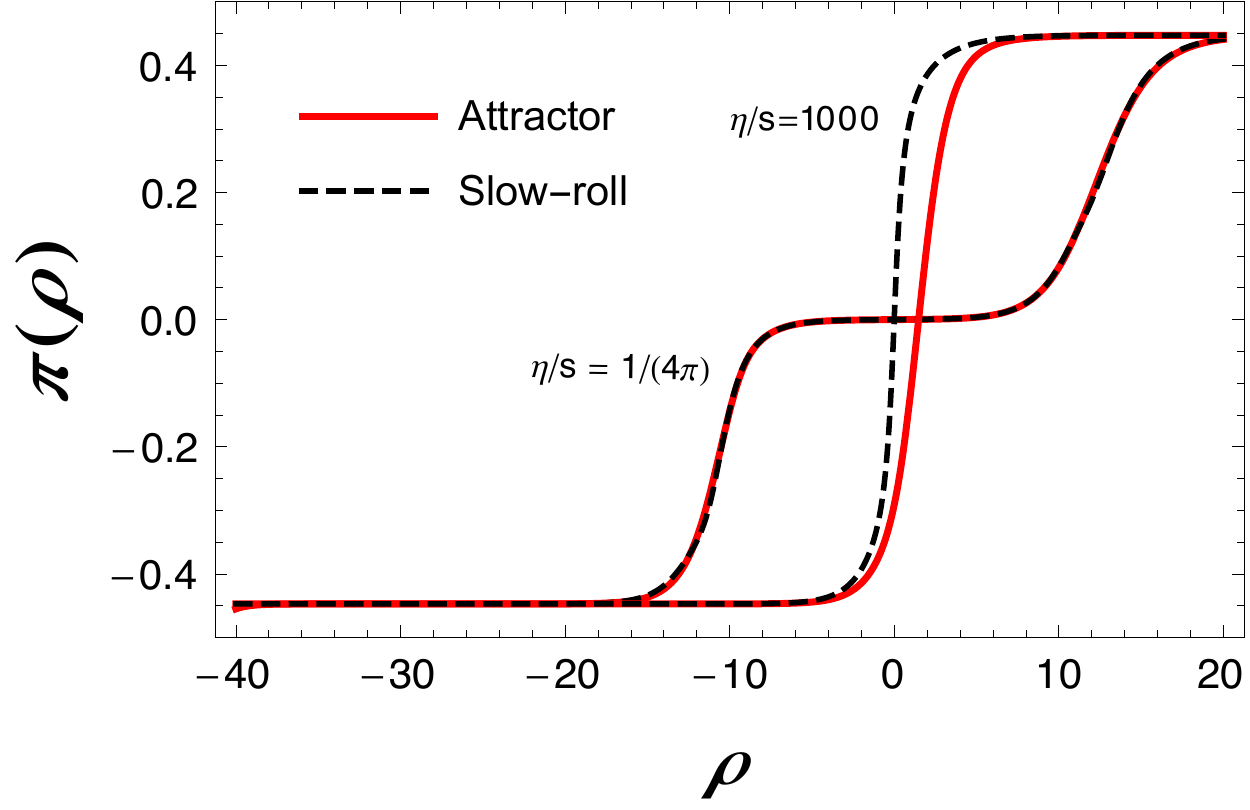}
\caption{(Color online) Comparison between the attractor solutions of IS
equations undergoing Gubser flow, for $\protect\eta/s=1/(4\protect\pi)$ and $%
\protect\eta/s=1000$, and the corresponding approximations using the
slow-roll series (dashed).}
\label{fig:pietascomparaattractor}
\end{figure}

Now we explore a different feature, so far exclusive to Gubser flow, which
is the fact that the attractor solution per se depends on the temperature.
This can be seen in Fig.\ \ref%
{fig:pietas14piTdependenceattractorfixedpicomparisonSR} where we now fix $%
\pi (-30)=-1/\sqrt{5}$ (attractor solutions) and consider three very
different values for the initial temperature $T(-30)=9.222\times 10^{-8}$, $%
T(-30)=9.222\times 10^{-10}$, and $T(-30)=9.222\times 10^{-12}$ and $\eta
/s=1/(4\pi )$. These variations in initial temperature lead to very
different values for the temperature at $\rho =0$ in the attractor
solutions. One obtains three different profiles for the attractor, which are
also compared to their corresponding slow-roll series truncated at 2nd order
for the first two solutions and at 0th order for the last one. Even though $%
\eta /s$ is small, by significantly decreasing the initial value of the
temperature one can again recover the large $\tau _{R}(0)$ regime where the
slow-roll series does not work well.

\begin{figure}[th]
\includegraphics[width=0.5%
\textwidth]{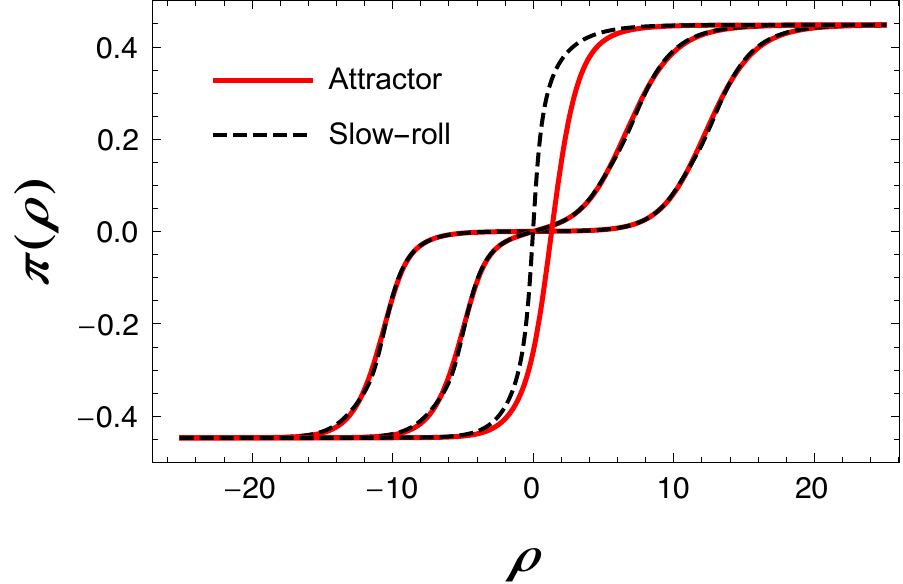}
\caption{(Color online) Comparison between the attractor solutions of IS
equations undergoing Gubser flow, computed using different initial
conditions for the temperature, and the corresponding approximations using
the slow-roll series (dashed). In this plot $\protect\eta/s=1/(4\protect\pi)$%
. }
\label{fig:pietas14piTdependenceattractorfixedpicomparisonSR}
\end{figure}

\begin{figure}[th]
\includegraphics[width=0.5\textwidth]{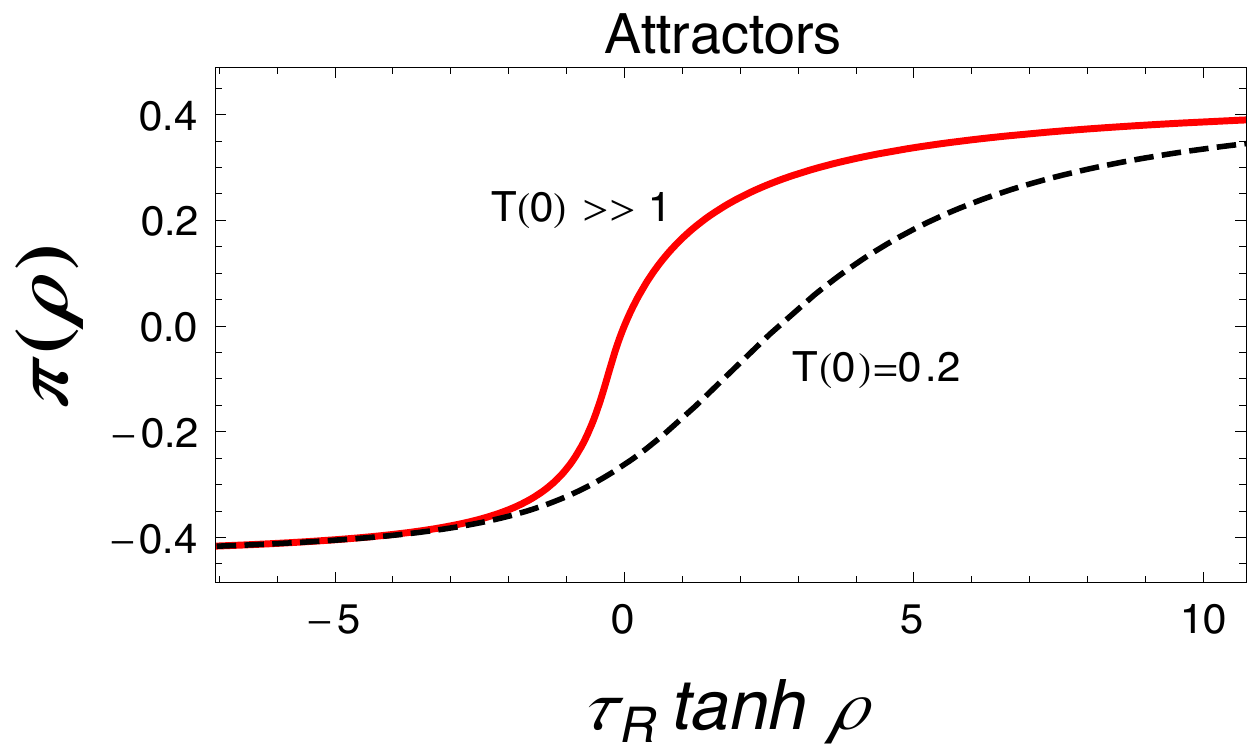}
\caption{(Color online) Attractor solutions of IS equations undergoing
Gubser flow, computed using different initial conditions for the
temperature, versus the Knudsen number-like quantity $\protect\tau_R \tanh%
\protect\rho$. The fact that these curves are distinct imply that in Gubser
flow the attractor solution of IS theory is not just a simple function of $%
\protect\tau_R \tanh\protect\rho$. Rather, the attractor depends on both $%
\protect\tau_R$ and $\tanh\protect\rho$, separately. In this plot $\protect%
\eta/s=1/(4\protect\pi)$. }
\label{fig:pietas14piattractorfock}
\end{figure}

Finally, in Fig.\ \ref{fig:pietas14piattractorfock} we explore another key
difference between the Gubser flow attractor and the Bjorken flow attractor.
In this figure we picked two of the attractor solutions discussed above
where $T(0)=0.2$ and $T(0)\gg 1$, with $\eta /s=1/(4\pi )$, and plotted them
against the Knudsen number combination $\tau _{R}\tanh \rho $. The fact that
these curves are different show that, in contrast to the Bjorken flow case,
the attractor solution of Israel-Stewart equations undergoing Gubser flow is
not solely a function of $\tau _{R}\tanh \rho $, even though the 0th order
truncation of the slow-roll series is. This illustrates that characterizing
attractor solutions by the 0th order slow-roll series can be misleading as
this special solution of Israel-Stewart equations (the attractor) displays a
more complex dependence on $\rho $ than the simplest truncated series. Since
already the 1st order truncation of the slow-roll series depends on both $%
\tau _{R}$ and $\tanh \rho $, we see that higher order truncations of the
slow-roll series are better suited to properly characterize the attractor.

\section{Conclusions}

\label{sec:conclusions}

In this paper we developed a perturbative scheme to construct asymptotic
solutions, such as the gradient and slow-roll expansions, of Israel-Stewart
theory undergoing Gubser flow. We then determine for the first time the large
order behavior of these perturbative expansions in the case of Gubser flow.
We demonstrated numerically that the expansion coefficients in both cases
grow factorially, indicating that these series have a zero radius of
convergence. Even though these series appear to diverge, their low order
truncations can still offer a reasonable description of exact solutions of
Israel-Stewart theory near the origin of the Gubser coordinate system.
However, we emphasize that such agreement is only possible when the
relaxation time is sufficiently small at $\rho =0$, i.e., $\tau _{R}\left(
\rho =0\right) \ll 1$.

When comparing both asymptotic solutions, we found that the slow-roll
expansion provides a much better overall description of exact solutions of
Israel-Stewart, when truncated at low orders. In particular, a truncated
slow-roll expansion can even describe qualitative and, often, quantitative,
aspects of exact solutions when $\rho \rightarrow \infty $ -- a region in
which gradients cannot be considered small. This suggests the existence of
nontrivial constitutive relations satisfied by the shear stress tensor that 
are valid even when gradients are large (i.e., the far-from-equilibrium regime).

We also demonstrated that the slow-roll series converges to $\pm 1/\sqrt{5}$
when $\left\vert \rho \right\vert \rightarrow \infty $ and used this fact to
numerically construct a resummed version of the slow-roll expansion. We
showed numerically that such solution displays the basic properties expected
of an attractor, with solutions obtained using different initial conditions
always converging to such resummed solution. Differently than the case of
Bjorken flow, we found that the Gubser flow attractor solution cannot be
expressed just as a function of the effective Knudsen number $\sim \tau _{R}\,\tanh \rho $. Instead,
it depends on the relaxation time and $\tanh \rho $ separately, suggesting
the existence of a class of attractor solutions and not just a single
universal function. Nevertheless, we emphasize that when the relaxation time
is sufficiently small at $\rho =0$, $\tau _{R}\left( 0\right) \ll 1$, the
attractor can be approximately expressed as a function of $\tau _{R}\,\tanh
\rho $. Furthermore, the truncated slow-roll series can also describe the
attractor solution of Gubser flow as long as the system is sufficiently
close to equilibrium near the origin (i.e., $\rho =0$).

Finally, our results give further support to the idea that new \emph{resummed}
constitutive relations between dissipative currents and the gradients of
conserved quantities can emerge in systems far from equilibrium that are
beyond the regime of validity of the usual gradient expansion.

\section*{Acknowledgements}

The authors than R.~Critelli for discussions about the numerical work
performed in this paper. JN and GSD thank Conselho Nacional de
Desenvolvimento Cient\'{\i}fico e Tecnol\'{o}gico (CNPq) for financial
support. JN thanks Funda\c{c}\~{a}o de Amparo \`{a} Pesquisa do Estado de S%
\~{a}o Paulo (FAPESP) under grant 2015/50266-2 for financial support.

\appendix

\section{Yet another implementation of the gradient expansion}

\label{sec:appendixA}

In this appendix we consider a possible solution for $T$ and $\pi $ that is
represented as a series in powers of $\epsilon $ 
\begin{eqnarray}
T &\sim &\sum\limits_{n=0}^{\infty }T_{n}\left( \rho \right) \epsilon ^{n},
\\
\pi &\sim &\sum\limits_{n=0}^{\infty }\pi _{n}\left( \rho \right) \epsilon
^{n}.
\end{eqnarray}%
Substituting the proposed series solutions in powers of $\epsilon $ into the
equations of motion for $\pi $ and $T$ one obtains the following set of 
\textit{coupled} equations:%
\begin{eqnarray}
\sum_{n=0}^{\infty }\partial _{\rho }T_{n}\epsilon ^{n}+\frac{2}{3}%
\sum_{n=0}^{\infty }T_{n}\epsilon ^{n}\tanh \rho -\frac{1}{3}%
\sum_{n=0}^{\infty }\sum_{m=0}^{\infty }\pi _{n}T_{m}\epsilon ^{n+m}\tanh
\rho &=&0, \\
c\sum_{n=0}^{\infty }\partial _{\rho }\pi _{n}\epsilon
^{n+1}+\sum_{n=0}^{\infty }\sum_{m=0}^{\infty }\pi _{n}T_{m}\epsilon ^{n+m}+%
\frac{4}{3}c\sum_{n=0}^{\infty }\sum_{m=0}^{\infty }\pi _{n}\pi _{m}\epsilon
^{n+m+1}\tanh \rho &=&\frac{4}{15}c\epsilon \tanh \rho .
\end{eqnarray}%
We now group together the terms that are of the same power in $\epsilon $,
obtaining the set of recurrence relations that must be solved to obtain $%
T_{n}$ and $\pi _{n}$. The terms that are of zeroth order in $\epsilon $
satisfy%
\begin{eqnarray}
\partial _{\rho }T_{0}+\frac{2}{3}T_{0}\tanh \rho -\frac{1}{3}\pi
_{0}T_{0}\tanh \rho &=&0, \\
\pi _{0}T_{0} &=&0,
\end{eqnarray}%
leading to the solution%
\begin{eqnarray}
\partial _{\rho }T_{0}+\frac{2}{3}T_{0}\tanh \rho &=&0, \\
\pi _{0} &=&0.
\end{eqnarray}%
Note that the equations above are exactly the same as those of an ideal
fluid and, consequently, the lowest order truncation of the series leads to
ideal hydrodynamics, as expected.

Collecting the terms that are of first order in $\epsilon $, one obtains%
\begin{eqnarray}
\partial _{\rho }T_{1}+\frac{2}{3}T_{1}\tanh \rho -\frac{1}{3}\pi
_{1}T_{0}\tanh \rho &=&0, \\
\pi _{1} &=&\frac{4c}{15T_{0}}\tanh \rho ,
\end{eqnarray}%
leading to the following equation for $T_{1}$%
\begin{equation}
\partial _{\rho }T_{1}+\frac{2}{3}T_{1}\tanh \rho =\frac{4c}{45}\tanh
^{2}\rho .
\end{equation}%
Furthermore, the terms that are of second order or higher in $\epsilon $, ($%
n\geq 2$), satisfy the equations 
\begin{eqnarray}
\partial _{\rho }T_{n}+\frac{2}{3}T_{n}\tanh \rho -\frac{1}{3}%
\sum_{m=0}^{n}T_{n-m}\pi _{m}\tanh \rho &=&0, \\
c\partial _{\rho }\pi _{n}+T_{0}\pi _{n+1}+\sum_{m=1}^{n}T_{m}\pi _{n-m+1}+%
\frac{4c}{3}\sum_{m=1}^{n}\pi _{n-m}\pi _{m}\tanh \rho &=&0.
\end{eqnarray}

The equations/solutions obtained above have a rather disturbing feature. So
far, it was not possible to derive a constitutive equation for $\pi $ that
is expressed solely in terms of $T$ and $\tanh \rho $ (gradients of
velocity), as would be expected in a gradient expansion. Instead, the
solution appears in terms of the temperature expansion coefficients, $T_{n}$%
. Nevertheless, we will later demonstrate that this solution, truncated up
to a given order in $\epsilon $, can be resumed and re-expressed solely in
terms of the full temperature $T$. For the first order truncation, this
procedure is rather obvious: one simply takes the $1/T_{0}$ dependence, that
appears in the solution for $\pi _{1}$, and re-expresses it as: $1/T_{0}\sim
1/T+\mathcal{O}(\epsilon )$. Therefore, neglecting terms of second order or
higher in $\epsilon $, we have that%
\begin{equation}
\pi =\pi _{0}+\epsilon \pi _{1}+\mathcal{O}(\epsilon ^{2})=\frac{4c\epsilon 
}{15T}\tanh \rho +\mathcal{O}(\epsilon ^{2}).
\end{equation}%
Similarly, the equation of motion for the temperature up to second order, $%
T_{0}+\epsilon T_{1}$, can be written as%
\begin{equation}
\partial _{\rho }\left( T_{0}+\epsilon T_{1}\right) +\frac{2}{3}\left(
T_{0}+\epsilon T_{1}\right) \tanh \rho =\frac{4c}{45}\epsilon \tanh ^{2}\rho
,
\end{equation}%
and, consequently, we have that%
\begin{equation}
\partial _{\rho }T+\frac{2}{3}T\tanh \rho =\frac{4c}{45}\epsilon \tanh
^{2}\rho +\mathcal{O}(\epsilon ^{2}).
\end{equation}%
If we set $\epsilon =1$, the equation above becomes the Navier-Stokes
equation under Gubser flow. This equation can be solved analytically, as was
shown in Ref.\ \cite{Gubser}, even though this solution displays unphysical
features such as negative temperatures when $\rho \to -\infty$.

Next, we obtain the second and third order solutions and show that these can
also be expressed in terms of the full temperature (up to the corresponding
order). The second order equations (for $T_{2}$ and $\pi _{2}$) are%
\begin{eqnarray}
\partial _{\rho }T_{2}+\frac{2}{3}T_{2}\tanh \rho -\frac{1}{3}\pi
_{2}T_{0}\tanh \rho -\frac{1}{3}\pi _{1}T_{1}\tanh \rho &=&0, \\
c\partial _{\rho }\pi _{1}+\pi _{2}T_{0}+\pi _{1}T_{1} &=&0.
\end{eqnarray}%
Using the solutions/equations already obtained for $\pi _{1}$, $T_{1}$, and $%
T_{0}$, we find the constitutive equation satisfied by $\pi _{2}$,%
\begin{equation}
\pi _{2}=\frac{4c^{2}}{15T_{0}^{2}}\left( -1-\frac{T_{1}}{c}\tanh \rho +%
\frac{1}{3}\tanh ^{2}\rho \right) .
\end{equation}%
One can see that $\pi _{2}$ depends separately on $T_{0}$ and $T_{1}$, while
the equation of motion for $T_{2}$ is coupled to $T_{0}$, $T_{1}$, $\pi _{1}$%
, and $\pi _{2}$.

The third order equations are%
\begin{eqnarray}
\partial _{\rho }T_{3}+\frac{2}{3}T_{3}\tanh \rho -\frac{1}{3}T_{2}\pi
_{1}\tanh \rho -\frac{1}{3}T_{1}\pi _{2}\tanh \rho -\frac{1}{3}T_{0}\pi
_{3}\tanh \rho &=&0, \\
c\partial _{\rho }\pi _{2}+\pi _{3}T_{0}+\pi _{2}T_{1}+\pi _{1}T_{2}+\frac{4%
}{3}c\pi _{1}^{2}\tanh \rho &=&0.
\end{eqnarray}%
Using all the equations obtained for the lower order coefficients, the
constitutive equation satisfied by $\pi _{3}$ can be simplified to 
\begin{eqnarray}
\pi _{3} &=&\frac{4c^{2}}{15T_{0}^{2}}\frac{T_{1}}{T_{0}}\left( 1+\frac{T_{1}%
}{c}\tanh \rho -\frac{1}{3}\tanh ^{2}\rho \right) -\frac{4c}{15T_{0}}\frac{%
T_{2}}{T_{0}}\tanh \rho \\
&&+\frac{4c^{3}}{15T_{0}^{3}}\left( \frac{T_{1}}{c}+\frac{2}{3}\tanh \rho -%
\frac{1}{3}\frac{T_{1}}{c}\tanh ^{2}\rho -\frac{2}{45}\tanh ^{3}\rho \right)
.
\end{eqnarray}%
The second and third order coefficients, $\pi _{2}$ and $\pi _{3}$, are
complicated and contain terms that have mixed contributions from $T_{0}$, $%
T_{1}$, and $T_{2}$.

We shall now re-express all these contributions solely in terms of $T$. In
order to perform this task, one should first note that%
\begin{eqnarray}
\pi _{1} &=&\frac{4\tau _{R}}{15}\left( 1+\epsilon \frac{T_{1}}{T}+\epsilon
^{2}\frac{T_{2}}{T}+\epsilon ^{2}\frac{T_{1}^{2}}{T^{2}}\right) \tanh \rho +%
\mathcal{O}\left( \epsilon ^{3}\right) , \\
\pi _{2} &=&-\frac{4}{15}\tau _{R}^{2}\left( 1-\frac{1}{3}\tanh ^{2}\rho +%
\frac{T_{1}}{c}\tanh \rho \right) -\frac{8T_{1}}{15T}\epsilon \tau
_{R}^{2}\left( 1-\frac{1}{3}\tanh ^{2}\rho +\frac{T_{1}}{c}\tanh \rho
\right) +\mathcal{O}\left( \epsilon ^{2}\right) , \\
\pi _{3} &=&\frac{4\tau _{R}^{3}}{15}\left[ \frac{2}{3}\tanh \rho -\frac{2}{%
45}\tanh ^{3}\rho +\frac{T_{1}}{c}\left( 1-\frac{1}{3}\tanh ^{2}\rho \right) %
\right] -\frac{4\tau _{R}}{15}\frac{T_{2}}{T}\tanh \rho \\
&&+\frac{4\tau _{R}^{2}}{15}\left( 1-\frac{1}{3}\tanh ^{2}\rho +\frac{T_{1}}{%
c}\tanh \rho \right) \frac{T_{1}}{T}+\mathcal{O}\left( \epsilon \right) .
\end{eqnarray}%
All that was done above was to rewrite $T_{0}$ in terms of $T$, up to a
given order in $\epsilon $. Combining all these expressions, $\pi =\epsilon
\pi _{1}+\epsilon ^{2}\pi _{2}+\epsilon ^{3}\pi _{3}+\mathcal{O}\left(
\epsilon ^{4}\right) $, one can verify that all contributions including $%
T_{1}$ and $T_{2}$ cancel each other and only the dependence on the full
temperature is left. The resumed answer is 
\begin{eqnarray}
\pi &=&\frac{4}{15}\epsilon \tau _{R}\tanh \rho -\frac{4}{15}\left( \epsilon
\tau _{R}\right) ^{2}\left( 1-\frac{1}{3}\tanh ^{2}\rho \right) \\
&&+\frac{8}{45}\left( \epsilon \tau _{R}\right) ^{3}\left( \tanh \rho -\frac{%
1}{15}\tanh ^{3}\rho \right) +\mathcal{O}\left( \epsilon ^{4}\right) .
\end{eqnarray}%
Therefore, we are able to rewrite the previous series as a power series in $%
\epsilon \tau _{R}$, with all the temperature contributions contained in the
powers of the relaxation time. Note that the equation of motion for the
(full) temperature can also be written in the form%
\begin{equation}
\partial _{\rho }T+\frac{2}{3}T\tanh \rho =\frac{1}{3}\pi T\tanh \rho +%
\mathcal{O}(\epsilon ^{4}).
\end{equation}

Finally, following this scheme we obtain equations of motion for the
temperature that are complemented by constitutive equations for the shear
stress tensor. These equations have the form that is traditionally
associated with a gradient expansion, as discussed in the main text.


\end{document}